\documentclass[11pt]{article}

\usepackage[left=1.5cm, right=1.5cm, top=3.0cm, bottom=3.0cm]{geometry}
\usepackage{graphicx,color}
\usepackage{array,tabularx,booktabs,geometry,subfigure,graphicx,longtable}
\usepackage{amsmath,amssymb,amsfonts}
\usepackage{extarrows}
\usepackage[hyperfootnotes=false, linktocpage=true, colorlinks, citecolor=blue, linkcolor=blue, urlcolor=blue]{hyperref}
\usepackage[all]{xy} 
\usepackage[mathscr]{eucal}
\usepackage{hypbmsec}
\usepackage{fancyvrb}
\usepackage{flexisym}
\usepackage{float}
\usepackage{fixltx2e,mathtools}
\usepackage{cite}
\usepackage[bf]{caption}
\def\bZ{\mathbb{Z}}

\def\bC{\mathbb{C}}
\def\bP{\mathbb{P}}

\def\cA{\mathcal{A}}
\def\cN{\mathcal{N}}
\def\cO{\mathcal{O}}

\def\cL{\mathcal{L}}

\def\tr{\mathop{\mathrm{tr}}}

\DeclareMathOperator{\Vol}{Vol}
\DeclareMathOperator{\dVol}{dVol}

\def\fnote#1#2{\begingroup\def\thefootnote{#1}\footnote{#2}
     \addtocounter{footnote}{-1}\endgroup}

\begin{document}

\title{ {\Large \bf Numerical Hermitian Yang-Mills Connection \\ for Bundles on Quotient Manifold}}

\author{
Wei~Cui${}{}$
}
\date{}
\maketitle
\begin{center} {\small 
${}$
{\it Yanqi Lake Beijing Institute of Mathematical Sciences and Applications (BIMSA),\\ Huairou District, Beijing 101408, China}
\vspace{0.4cm} \\ 
${}$
{\it Yau Mathematical Sciences Center, Tsinghua University, Beijing 100084, China} }


\fnote{}{cwei@bimsa.cn}

\end{center}

\begin{abstract}
\noindent

We extend the previous computations of Hermitian Yang-Mills connections for bundles on complete intersection Calabi-Yau manifolds to bundles on their free quotients. 
Bundles on quotient manifolds are often defined by equivariant bundles on corresponding covering spaces. 
Combining equivariant structure and generalized Donaldson's algorithm, we develop a systematic approach to compute connections of holomorphic  poly-stable bundles on quotient manifolds. 
With it, we construct the connections of an $SU(3)$ bundle on $Z_5$ quotient of quintic 
and an $SU(5)$ bundle on $Z_3 \times Z_3$ quotient of Bi-cubic.
For both of these examples, the algorithm converges as expected and gives a good approximation of Hermitian Yang-Mills connections, which will be important for heterotic model building in Calabi-Yau compactification.

\end{abstract}

\thispagestyle{empty}
\setcounter{page}{0}
\newpage

\tableofcontents

\section{Introduction} \label{sec:sec1}

How to reproduce the real world physics, in particular the Standard Model of particle physics, is a fundamental question in string theory. 
One promising approach is the compactification of the $E_8 \times E_8$ heterotic strings on a Calabi-Yau (CY) threefold $X$ with a holomorphic, poly-stable bundle $V$~\cite{Candelas:1985en}. 
The $\cN=1$ supersymmetry requires that the metric on $X$ to be the {\it Ricci-flat metric} (RFM)~\cite{Yau1978OnTR}, and the connection of $V$ should satisfy the Hermitian Yang-Mills (HYM) equation~\cite{Green:1987mn}
\begin{equation} \label{eqn:HYMeqn}
  F_{a b}=F_{\bar{a}\bar{b}}=0, \qquad  
  g^{\bar{b}a}F_{a\bar{b}}=0,  
\end{equation} 
where $g^{\bar{b}a}$ is the RFM of $X$, $F_{a\bar{b}}$ is the field strength of the connection $A_{a}$ and $a,\bar b=1,2,3$.
The connection with this property is called {\it HYM connection} and it exists uniquely for poly-stable bundles~\cite{Donaldson:1985zz,UYau}.

Given such a vacua, the 4d low-energy effective theory is obtained by a two-step process. First, the dimensional reduction on the CY manifold leads to intermediate grand unified theories (GUTs) typically with gauge group $SU(5)$ or $SO(10)$. Then, one can turn on Wilson lines in the vacuum, which could break the gauge group down to $SU(3) \times SU(2) \times U(1)$ and project out many unwanted matter components~\footnote{The other advantage is the non-simply connected threefolds have many fewer moduli to stabilize \cite{Anderson:2010mh,Anderson:2011ty,  Anderson:2011cza,Anderson:2013qca,Anderson:2009sw,Anderson:2009nt,Cui:2020ijv} compared to their covering spaces.}. 
In this way, many heterotic models are constructed whose charged matters are identical to the spectrum of the Minimal Supersymmetric Standard Model (MSSM)~\cite{Greene:1986bm, Greene:1986jb,Bouchard:2005ag,Braun:2005zv,Braun:2005bw,Braun:2005ux,Anderson:2009mh,Braun:2011ni,Anderson:2011ns,Anderson:2012yf,Anderson:2013xka,Gray:2019tzn}. 
Note that the Wilson lines used in the second step only exist on manifolds with $\pi_1(X) \neq 0$. 
Thus, to obtain realistic models, we need to work with the non-simply connected CY manifolds.

Besides the spectrum, one would like also to determine the Yukawa couplings in these models. 
In the 4d effective theories, it is known in \cite{Strominger:1985ks} that holomorphic Yukawa couplings from the superpotential is quasi-topological and they can be computed algebraically \cite{Distler:1987ee,Candelas:1987se,Candelas:1987rx,Distler:1995bc,Braun:2006me,Bouchard:2006dn,Anderson:2009ge,Anderson:2010tc,Blesneag:2015pvz,Blesneag:2016yag,Anderson:2021unr}.
But they are not physical. To obtain the physical couplings, one needs to  normalize the holomorphic ones by metric from K\"ahler potential.
However, the K\"ahler potential depends on the explicit form of the RFM on $X$ and the HYM connection of $V$ \cite{Blesneag:2018ygh}, which are notably difficult to compute.

In recent years, several numerical methods are developed to solve this problem for example the (generalized) Donaldson algorithm \cite{donaldson_2009, Douglas:2006rr,Braun:2007sn,Douglas:2006hz,Anderson:2010ke,Anderson:2011ed}, minimization algorithm \cite{Headrick:2009jz, Headrick:2005ch, Cui:2019uhy} and more recently machine learning algorithms \cite{Ashmore:2019wzb, Douglas:2020hpv,
Anderson:2020hux, Jejjala:2020wcc,Ashmore:2021rlc,Larfors:2021pbb,Larfors:2022nep,Ashmore:2021ohf,Berglund:2022gvm,Gerdes:2022nzr}. 
By them, one can compute the RFM of a CY manifold and the HYM connection of a poly-stable bundle to a high degree of accuracy. 
However, these numerical methods have only been applied to simply connected CY manifolds. 
It is important for us to extend them to non-simply ones since, as we discussed above, those are CY manifolds that leads to realistic heterotic models.

The non-simply connected CY manifold can often be realized as free quotient of the simply connected CY manifold by discrete symmetry $\Gamma$ \cite{Braun:2010vc,Braun:2017juz, Gray:2021kax}. 
The RFM on the quotient CY manifold has been computed using Donaldson's algorithm in \cite{Braun:2007sn}. 
In this paper, we will study how to use generalized Donaldson's algorithm to compute the HYM connection of a holomorphic, poly-stable bundle on the quotient manifolds.

The holomorphic bundles on the quotient manifold, sometimes called ``downstairs bundles", can be defined by bundles on the covering spaces, i.e. ``upstairs bundles", that admits $\Gamma$-equivariant structures \cite{Anderson:2009mh,Braun:2009mb,Anderson:2012yf}. 
For globally generated upstairs bundles, the equivariant structure is specified by a set of sections-wise maps, which depends on the $\Gamma$-action on the global sections and a choice of bundle morphism  \cite{Anderson:2009mh}. 
The global sections of the downstairs bundle are thus given by the $\Gamma$-invariant sections of the corresponding upstairs bundle, with which we can apply the generalized Donaldson's algorithm. 

To illustrate our method, we consider an $SU(3)$ monad bundle on a $Z_5$ quotient of the quintic threefold and an $SU(5)$ monad bundle on a $Z_3 \times Z_3$ quotient of Bi-cubic threefold. The corresponding upstairs bundles have equivariant structure involving both trivial and non-trivial bundle morphism. 
We calculate the RFM of the $Z_5$ quotient of the quintic and the $Z_3 \times Z_3$ quotient of Bi-cubic and then apply the generalized Donaldson's algorithm to both monad bundles. Our results show that the algorithm converges as expected and provides good approximations of HYM connections for bundles on the quotient manifolds.

The paper is organized as follows. In section~\ref{sec:sec2}, we first review the equivariant structure of holomorphic bundles and present the way to construct the invariant sections. After that, we study how to use Donaldson's algorithm and its generalization to compute the RFM of the quotient CY manifold and the HYM connection of the poly-stable bundle on it. The algorithm to perform numerical integration is introduced in the end.
In section~\ref{sec:sec3}, we apply the generalized Donaldson's algorithm just discussed to compute the HYM connection of an $SU(3)$ monad bundle on a $Z_5$ quotient of the quintic threefold and an $SU(5)$ monad bundle on a $Z_3 \times Z_3$ quotient of Bi-cubic. Then, we evaluate the performance and accuracy of the algorithm.
In section~\ref{sec:sec4}, we conclude our work and point out a few future directions.

\section{Generalized Donaldson's algorithm for bundles on quotient manifolds} \label{sec:sec2}

We will first review the construction of holomorphic bundles on the quotient manifolds and then study how to use the generalized Donaldson's algorithm to compute the HYM connection of the downstairs bundles. 
The Monte-Carlo algorithm for numerical integration is introduced in the end.

\subsection{Equivariant bundles} \label{sec:subsec21}

To obtain realistic four-dimensional models, 
one needs to turn on the Wilson lines in the vacua, which requires the internal CY threefold to be non-simply connected~\cite{Greene:1986bm, Greene:1986jb,Bouchard:2005ag,Braun:2005zv,Braun:2005bw,Braun:2005ux,Anderson:2009mh,Braun:2011ni,Anderson:2011ns,Anderson:2012yf,Anderson:2013xka,Gray:2019tzn}. 
However, most of known CY threefolds are simply connected, such as the complete intersection CY manifolds~\cite{Candelas:1987kf,Anderson:2015iia} and the Kreuzer-Skarke CY manifolds~\cite{Kreuzer:2000xy}. 
To obtain non-simply connected ones, let's consider a simply connected CY manifold with discrete symmetry $\Gamma$. When the symmetry is fix-points free~
\footnote{
In general, checking if the $\Gamma$-action is free is nontrivial. Interested reader are referred to \cite{Braun:2010vc,Braun:2017juz, Gray:2021kax} for more details. }, one can show that the quotient manifold $\hat{X} = X/\Gamma$ has $\pi_{1}(\hat{X})\neq 0$. 
In this way, given any CY threefolds admitting free-acting symmetry, one can construct non-simply connected CY manifolds.

The other object in a heterotic vacua is a holomorphic downstairs bundle $\hat{V}$ on $\hat{X}$, which can be often defined by a $\Gamma$-equivariant upstairs bundle $V$ on the covering space~\cite{Anderson:2009mh,Braun:2009mb,Anderson:2012yf}.
A holomorphic vector bundle $V \stackrel{\pi}{\rightarrow} X$ is $\Gamma$-equivariant if it admits a set of bundle morphism $\phi_g: V
\rightarrow V$ for each $g \in \Gamma$, such that they satisfying the following two condition. 
First, each $\phi_g$ commute with the projection map $\pi$, i.e. $\phi_g \circ \pi = \pi \circ \phi_g$ and covers the action $g:X\rightarrow X$ on the base, which is equivalent to require the diagram below 
\begin{equation*}
  \begin{array}{lllll}
  &V&\stackrel{\phi_g}{\longrightarrow}&V&\\
  \pi &\downarrow&&\downarrow&\pi\\
  &X&\stackrel{g}{\longrightarrow}&X&
 \end{array}
\end{equation*} 
commutes for all $g \in \Gamma$. 
Second, the set of morphisms should obey the co-cycle
condition, i.e. for all $g,h \in \Gamma$, 
\begin{equation*}
\phi_g \circ \phi_h =\phi_{gh} \;. 
\end{equation*}
If such a set of morphisms exist, $V$ has a equivariant structure and descends to a bundle $\hat{V}$ on $\hat{X}$. In the rest of the paper, we will mainly work with the equivariant bundle $V$ because in principle, all the quantities of $\hat{V}$, such as  Chern classes, cohomologies \cite{Anderson:2009mh}, and even connection can be computed from the corresponding upstairs bundle $V$.

Suppose $V$ is an $\Gamma$-equivariant bundle on $X$. The bundle morphism $\phi_g$, for each $g\in \Gamma$, induces an action on global section $s \in H^0(X,V)$ by the following commutative diagram,
\begin{equation*}
  \begin{array}{lllll}
  &V&\stackrel{\phi_g}{\longrightarrow}&V&\\
  s &\uparrow&&\uparrow&s'\\
  &X&\stackrel{g}{\longrightarrow}&X&
 \end{array} \;.
 \label{invalts}
\end{equation*}
where the {\it section-wise maps}  
$\Phi_g$ are defined by  
\begin{equation}\label{eqn:InvSection}
 s'=\Phi_g(s)=\phi_g\circ s\circ g^{-1}\; .
\end{equation} 
Obviously, these actions also satisfy the condition
\begin{equation}\label{eqn:cocyleC}
    \Phi_g\circ \Phi_h=\Phi_{gh}\;, \qquad g,h \in \Gamma
\end{equation}
Thus, for a $\Gamma$-equivariant vector bundle, there exist a set of maps $\Phi_g$ between $H^0(X, V)$ with $g\in \Gamma$ such that they form a representations of $\Gamma$.

In general, for a given holomorphic bundle, checking if it is $\Gamma$-equivariant by constructing the bundle morphisms $\{\phi_g\}$ explicitly is difficult.  
To simplify the analysis, we will consider the bundles that are global generated. 
Instead of the bundle morphisms, for a globally generated bundle $V$, the $\Gamma$-equivariant structure can also be constructed using the section-wise maps $\{\Phi_g\}$ between $H^0(X,V)$, most of time, that is a vector space spanned by a set of homogeneous polynomials which is much easier to work with.
From the equation (\ref{eqn:InvSection}), the maps $\{\Phi_g\}$ are obtained by a $\Gamma$-action on the homogeneous polynomials, i.e. $s\rightarrow s\circ g^{-1}$, and then composed with a bundle morphism $\phi_g$. $V$ is an equivariant bundle if there exist bundle morphisms such that the set of maps $\{\Phi_g\}$ is a representation of $\Gamma$.

Next, let's discuss how to construct equivariant vector bundles. 
All the bundles considered in this work will be built from the line bundles via the operation like tensor product, direct sum and the dual. It is easy to check that 
the equivariant structure are compatible with these operations. 
Suppose $V_1$ and $V_2$ are two equivariant bundles, then $V_1 \otimes V_2$, $V_1 \oplus V_2$ and $V^{*}_1$ or $V^*_2$ are also equivariant. However, the reverse statement is not necessary true. As we will see later, the direct sum of non-equivariant line bundles could be equivariant by a suitable choice of bundle morphisms $\{\phi_g\}$. We will see an example of that in section \ref{sec:sec3}.

Besides them, we can also define the higher-rank bundle $V$ by the monad construction below
 \cite{Okonek1980,Distler:1987ee,Hubsch:1992nu}
\begin{equation}  \label{eqn:monad}
0\rightarrow V \stackrel{}{\rightarrow}
B\stackrel{f}{\rightarrow}C\rightarrow 0\; ,
\end{equation} 
where $B$ and $C$ be sums of line bundles over $X$ and $f \in H^0(X,B^*\otimes C)$ is a bundle morphism between them. 
According to \cite{Anderson:2009mh}, the bundle $V$ is equivariant if the following two conditions are satisfied. First, $B$ and $C$ are equivariant. In addition, $f$ should be an invariant section with respective to the equivariant structure induced by the ones of $B$ and $C$.

Global sections of the downstairs bundle $\hat{V}$ are given by the {\it $\Gamma$-invariant sections} of the corresponding equivariant upstairs bundle $V$ \cite{Donagi:2004su,Donagi:2004ub}. For a given section $s \in H^0(X,V)$, it is $\Gamma$-invariant if 
\begin{equation} \label{eqn:invSection2}
    s=\Phi_{V,g(s)}=\phi_{V,g}\circ s\circ g^{-1}\; , \quad \forall g \in \Gamma \;,
\end{equation}
where $\Phi_{V,g(s)}$ are the section-wise maps defined in (\ref{eqn:InvSection}). 
Denote $H^0(\hat{X},\hat{V})$ to be the space of these invariant sections of $V$. Its dimension $h^0(\hat{X},\hat{V})$ can sometimes be calculated from the index of $V$. Recall that the index of a holomorphic bundle $V$ is defined by 
\begin{equation*}
{\rm ind} (V) = \sum \limits_{i=0}^3 (-1)^i h^i(X,V) 
\
\end{equation*}
For bundles with only $h^0(X,V) \neq 0$, we have ${\rm ind} (V)$ $=$ $h^0(X,V)$. Thus, the number of independent global sections of $\hat{V}$ is 
\begin{equation} \label{eqn:indexG} 
    h^0(\hat{X},\hat{V})
    =  \frac{ h^0(X,V)}{|\Gamma|} \;.
\end{equation}
We will use this expression to check the invariant sections of equivariant bundles obtained using computer in section \ref{sec:sec3}.

\subsection{Generalized Donaldson's algorithm}\label{sec:subsec22}

In this section, we will introduce how to use (generalized) Donaldson's algorithm to compute RFM on quotient manifold $\hat{X}$ and HYM connection of holomorphic bundle $\hat{V}$ on it. 
Most of steps in the algorithm are the same as before. 
The only difference is that here, on the quotient manifold, we need to work with global sections of the downstairs bundle $\hat{V}$ which are invariant sections of the corresponding equivariant upstairs bundles $V$.

\subsubsection*{Donaldson's algorithm}

We will compute the RFM of the quotient manifold $\hat{X}$ using Donaldson's algorithm. To begin with, one needs to choose an equivariant ample line bundle $L$ on the covering space $X$ such that it descends to an ample line bundle $\hat{L}$ on $\hat{X}$. By definition, for large enough $k$, there exist non-trivial global sections of $\hat{L^k}$, which can be obtained from the invariant sections of $L^k$ by (\ref{eqn:invSection2}).

Given a basis of the global sections $s_{\alpha} \in H^0(\hat{X},\hat{L}^k)$ with $\alpha=1,2,\ldots,n_k$. 
One can write down an ansatz for the K\"ahler potential  
\begin{equation} \label{eqn:ansatz1}
  K_{h,k}
  =\frac{1}{k\pi} \ln 
  \sum_{\alpha,\bar{\beta}=0}^{n_{k}-1} 
  {\bar{s}}_{\bar{\beta}}
  h^{\bar{\beta}\alpha}_k
  s_{\alpha}, 
\end{equation}
where $h_k$ is an arbitrary constant Hermitian matrix. 
This ansatz gives a large number of metric on $\hat{X}$ parameterized by $h_k$. 
As shown in \cite{donaldson_2001,donaldson_2009}, for a given $k$, the RFM can be approximated by one of these metric called {\it balanced metric}, and as $k\to \infty$, the corresponding balanced metric will approach to the exact RFM. 
This result provide a systematic way to compute the RFM. In practice, the balance metric for first a few $k$ is a good approximation of RFM.

Now, let's briefly review how to compute the balanced metric. 
For each twist $k$, we denote the parameter in (\ref{eqn:ansatz1}) associated with the balanced metric to be $h_k^b$.
To find the parameters $h^b_k$ corresponding to the balanced metric, we define the $T$ operator
\begin{equation} \label{eqn:toper1}
  T(h)_{\alpha\bar{\beta}}=\frac{n_{k}}{\Vol_{CY}}
  \int_{\hat{X}}
  \frac{s_{\alpha}{\bar{s}}_{\bar{\beta}}}{\sum_{\gamma\bar{\delta}}
    h^{\gamma\bar{\delta}}s_{\gamma}{\bar{s}}_{\bar{\delta}}}\dVol\;.
\end{equation} 
Here $ \dVol = \Omega \wedge \bar{\Omega}$ is the volume form of $\hat{X}$ defined by the nowhere-vanishing holomorphic $(3,0)$-form $\Omega$ and its complex conjugate \cite{Candelas:1987kf}, and the integration of it gives the volume 
\begin{equation*}
    \Vol_{CY} = \int_{\hat{X}} \Omega \wedge \bar{\Omega} \;. 
\end{equation*}
Notice that $h^b_k$ is the fixed point of $T$ operator. It can be shown \cite{donaldson_2001,donaldson_2009} that iterating the $T$ operator,
\begin{equation} \label{hiter1}
h^{i+1}_{k}=T(h^i_{k}),
\end{equation} 
starting from an arbitrary Hermitian matrix $h^{0}_{k}$, leads to a convergence towards the parameters $h^b_k$. 
Therefore, if we find fixed points under the iteration of the $T$ operator for high enough $k$, we should obtain a good numerical approximation to the RFM on $\hat{X}$. 
To summarize, Donaldson's algorithm is a reliable way to approximate the RFM of CY manifolds.



For a given matrix of parameters $h_k$, 
This can be used to define a top form $\omega_k^d$ on $\hat{X}$. With it, we can define an error function \cite{donaldson_2009}
\begin{equation} \label{eqn:error1}
\sigma_k=\frac{1}{\Vol_{CY}} \int_{\hat{X}} \left| 1-\frac{w^d_k/\Vol_k}{\Omega \wedge {\bar{\Omega}}/\Vol_{CY}}\right| \dVol, 
\end{equation}
where 
\begin{equation}
\label{volK}
    \Vol_k = \frac{1}{d!}\int_{\hat{X}} \omega_k^d
\end{equation}
is the volume associated with $\omega_k$. 
Clearly, this error function $\sigma_k$ vanishes if and only if $\omega_k$ is the K\"ahler form associated to the RFM. We will use this error function to evaluate the accuracy of our numerical metric.



\subsubsection*{Generalized Donaldson's algorithm}

Given a Calabi-Yau threefold $X$ with discrete action $\Gamma$, we have discussed how to use Donaldson's algorithm to compute the RFM on the quotient manifold $\hat{X}$. 
In this subsection, we will review the generalized Donaldson's algorithm and use it to compute the HYM connection of a poly-stable bundle $\hat{V}$ of rank $n$ on the quotient manifold $\hat{X}$ with the corresponding equivariant upstairs bundle denoted by $V$.

For a holomorphic bundle $\hat{V}$ with a hermitian fiber metric $G$, there exists an unique connection compatible with the complex structure and metric structure of $\hat{X}$ which is called {\it Chern connection} given by \cite{Griffiths:433962} 
\begin{equation} \label{eqn:chernConnection}
    A^{(1,0)}=(\partial G) G^{-1} \quad A^{(0,1)} =0 \;. 
\end{equation}
In terms of the fiber metric, the HYM equation becomes \cite{Douglas:2006hz,Anderson:2010ke,Anderson:2011ed}
\begin{equation} \label{eqn4:Hermitian Yang–Mills g1}
  g^{\bar{j}i} F_{i\bar{j}}= g^{\bar{j}i} \bar{\partial}_{\bar{j}}A_{i}
  =
  g^{\bar{j}i}\bar{\partial}_{\bar{j}}(G^{-1}\partial_i G) = 0
 .
\end{equation} 
The fiber metric is called {\it Hermite-Einstein} if it obeys the equation above and the corresponding Chern connection is exactly the HYM connection that we are looking for. We will compute this Hermite-Einstein fiber metric using the generalized Donaldson's algorithm.

Similar to the Donaldson's algorithm for the RFM, we will first choose an  ample equivariant line bundle $L$ on the covering space $X$, which descends to an ample line bundle $\hat{L}$ on $\hat{X}$. 
Since the bundle $\hat{V}$ is poly-stable, there are no global sections, i.e. $H^0(\hat{X},\hat{V})=0$. So, we will twist $\hat{V}$ with $\hat{L}^k$ for $k>1$, such that the twisted bundle $\hat{V}\otimes \hat{L}^k$ has non-trivial global sections. From the equation (\ref{eqn:invSection2}), these global sections can be obtained from the invariant sections 
of the equivariant upstairs bundle $V\otimes L^k$.

Let $\{S^i_{\alpha}\}$ be a basis of $H^0(\hat{X},\hat{V}\otimes \hat{L}^k)$ with $i=1,\cdots,n$ and $\alpha=1,\cdots, N_k=h^0(\hat{X},\hat{V}\otimes \hat{L}^k)$. 
Using these global sections, we have a ansatz for the fiber metric $\Tilde{G}$ of $\hat{V}\otimes \hat{L}^k$ as \cite{Douglas:2006hz,Anderson:2010ke,Anderson:2011ed}
\begin{equation}  \label{eqn:ansatz2}
  (\Tilde{G}^{-1})^{\bar{j}i} =
  \sum_{\alpha,\beta=0}^{N_k-1}
  (\bar{S})_{\bar{\beta}}^{\bar{j}}
  H^{\bar{\beta}\alpha}
  S_{\alpha}^{i},
\end{equation}
where $H=H_{\alpha\bar{\beta}}$ is a constant hermitian matrix with dimension $N_k$. Again, we will use the T-operator iteration to fix this $H$ matrix so that the fiber metric given by the equation (\ref{eqn:ansatz2}) is the {\it balanced fiber metric}. As shown in \cite{Wang,Seyyedali}, the balanced fiber metric of $\hat{V} \otimes \hat{L}^k$ after untwisting \cite{Anderson:2010ke} will converge to the Hermitian-Einstein fiber metric of $\hat{V}$ as $k \to \infty$.

To compute the balanced fiber metric, we define the T operator by 
\cite{Douglas:2006hz,Anderson:2010ke}
\begin{equation}\label{eqn:Toper2}
  T(H)_{\alpha\bar{\beta}}=
  \frac{N_{k}}{n \Vol_{CY}} \int_{\hat{X}}
  S^{i}_{\alpha}
  (\bar{S}^{\bar{j}}_{\bar{\delta}}
   H^{\bar{\delta}\gamma} 
   S^{i}_{\gamma})^{-1}
   \bar{S}^{\bar{j}}_{\bar{\beta}}
  \dVol \;. 
\end{equation}
The balanced fiber metric is determined by the fixed point of T-operator $T(H^b_k)=H^b_k$, which can be calculated by iterating the T-operator 
\begin{equation}\label{eqn:BalancedMetric2}
H^{i+1}_k = T(H^i_k)
\end{equation}
from an arbitrary initial matrix $H^{0}_{k}$. For most of cases, this iteration converges fast in about $50$ times.

Once we have computed the balanced fiber metric $\Tilde{G}$ of $\hat{V} \otimes \hat{L}^k$, the fiber metric of $\hat{V}$ can be obtained by a untwisting procedure using determinant line bundle \cite{Anderson:2010ke}. 
In terms of the fiber metric of the twisted bundle, the HYM equation (\ref{eqn:HYMeqn})  becomes  
\begin{equation} \label{eqn:untwistHermitian Yang-Mills }
g^{i \bar j} F_{i \bar j} = g^{i \bar j} \partial_{\bar j}(\Tilde{G}^{-1} \partial_i \Tilde{G}) - \frac{1}{n} \tr(g^{i \bar j} \partial_{\bar j}(\Tilde{G}^{-1} \partial_i \Tilde{G})) 1_{n \times n}\;.
\end{equation}
Here, $g^{i \bar j} F_{i \bar j}$ is a $n \times n$ matrix and will approach to zero matrix with $k \rightarrow \infty$. This allows us to define an error function \cite{Anderson:2010ke}
\begin{equation} \label{eqn:error2}
  \tau_k = 
  \frac{1}{n}
  \int_{\hat{X}} \Big( \sum |\lambda_i| \Big) 
  \dVol, 
\end{equation}
where $\lambda_i$, $i=1,\ldots, n$ are the eigenvalues of $g^{i \bar j} F_{i \bar j}$. Obviously, it approaches zero if the fiber metric after untwisting converges to Hermitian-Einstein one. We will use this error function to evaluate the accuracy of the balanced fiber metric computed using the Donaldson's algorithm.

\subsection{Numerical integration over Calabi-Yau threefold}\label{sec2:3}

In Donaldson's algorithm, the T-operator and the error function are defined as the integrals over the Calabi-Yau threefold. To compute them, we need to know how to evaluate these integrals. In this subsection, we will introduce the Monte-Carlo integration and use it to compute the integral over the CY manifolds. We will consider the complete intersection Calabi-Yau manifold (CICY). 
The Monte-Carlo integration over CICYs 
has been studied in \cite{Douglas:2006rr,Braun:2007sn,Douglas:2020hpv,Anderson:2020hux,Larfors:2021pbb}. We will closely follow them in this subsection \footnote{The Monte-Carlo integration over Kreuzer–Skarke Calabi–Yau manifolds is recently studied in \cite{Larfors:2022nep}.}. 
Note that algorithm to be described below also works for the integral over free quotients of CICYs.

The CICY is realized as the complete intersection of $K$ hypersurfaces inside the ambient space $\cA = \bP^{n_1} \times \bP^{n_2} \times \ldots \times  \bP^{n_m}$. The hypersurfaces are given by the defining homogeneous polynomials $p_i$ with degree ${q^j}_i$ where $i=1,2,\ldots, K$ and $j=1,2,\ldots, m$. In its most general form, such CY threefold can be expressed conveniently in the configuration matrix below
\begin{equation}\label{eqn:ConfMatCICY}
X = \def\arraystretch{1.2}
\left[
\begin{array}{c||ccc} 
    \bP^{n_1}&  q^{1}_{1} & \cdots &q^{1}_{K} \\
    \bP^{n_2} &  q^{2}_{1} & \cdots & q^{2}_{K}\\
    \vdots &  \vdots &\ddots&\vdots\\
    \bP^{n_m} & q^{r}_{1} & \cdots& q^{r}_{K}\\ 
\end{array}
\right]  \ .  
\end{equation}
where ${q^j}_i$ are chosen to obey the Calabi-Yau condition \cite{Candelas:1987kf}. We will only consider the simplest CICY where $X$ is just one hypersurface in the ambient space. But the method discussed in the following works for any CICY.

Consider a general integral over the CY threefold $X$
\begin{equation} \label{eqn:genIntegral}
    I=\int_X f \; \dVol \;,
\end{equation}
where $f$ is an arbitrary function defined on $X$ and $\dVol$ is the volume form. 
It is defined by the nowhere-vanishing holomorphic $(3,0)$-form $\Omega$ and its complex conjugate 
\begin{equation}
    \dVol = \Omega \wedge \bar\Omega \;.
\end{equation}
For the case of CICYs considered in this work, this $(3,0)$-form $\Omega$ can be determined analytically by the defining polynomials of $X$, see \cite{Candelas:1987kf} for details.

We will perform the definite integral $I$ using the Monte-Carlo methods \cite{James:1980yn}. 
The idea is that, instead of integrating over the entire domain $X$, the integral can be approximated by evaluating the integrand $f$ on a sample of finite points $p_i\in X$, where $i=1,\ldots ,N$. 
If these points are chosen according to some {\it auxiliary measure} $dA$, then the integral becomes 
\cite{Douglas:2006rr,Braun:2007sn} 
\begin{align}\label{eqn:intf1}
  \int_X f \dVol \; =\int_X f \,\frac{\dVol}{dA}\; dA \simeq\frac{1}{N}\sum_{i=1}^N  f(p_i)w_i\; ,
\end{align}
where $w_i = \dVol/dA$ is called {\it weights} of the point $p_i$. The statistical error of such an approximation is of order $1/\sqrt{N}$ times a quantity proportional
to $\langle f \rangle$, the mean of the integrand in the sample.

Different ways of choosing the points in the sample lead to different auxiliary measure $dA$. We will use the method proposed in \cite{Douglas:2006rr}, where the points are generated by the intersection of uniformly distributed random lines $L \subset \cA$ with the hypersurface $X$. In this setup, the auxiliary measure is expected to be \cite{Shiffman1999,Shiffman2008}
\begin{equation}
    dA = \langle L \cap X \rangle \sim i^*(\omega_{FS})
\end{equation}
where $i:X\to \cA$ is the restriction map and $\omega_{FS}$ is the Fubini-Study metric of the ambient space $\cA$. 
Given such a sample, the integral $I$ follows from the equation (\ref{eqn:intf1}). 
In the rest of the paper, all integral are performed using this Monte-Carlo method.

\section{Examples} \label{sec:sec3}

In this section, we will use the generalized Donaldson's algorithm to study the HYM connection of two concrete examples. In the first example, the equivariant structure of the upstairs bundle is defined with a trivial bundle morphism. While in the second example, we will consider a holomorphic bundle with equivariant structure involving non-trivial bundle morphism. As we will see that the numerical results converge to the HYM connection and the generalized Donaldson's algorithm can be applied to both of these examples.

\subsection{An $SU(3)$ monad bundle on a $Z_5$ quotient of Quintic}

The quintic CY threefold is a degree $5$ hypersurface in $\bP^4$ usually denoted by $X=[\bP^4|5]$. Let $z_i$, $i=0,\ldots 4$ be the homogeneous coordinates of $\bP^4$. 
We take the defining polynomial $P$ to the Fermat type as 
\begin{equation} \label{eqn:Fermat}
    P=z_0^5+z_1^5+z_2^5+z_3^5+z_4^5 \;. 
\end{equation}
There is a $\mathbb{Z}_5$ discrete symmetry 
\footnote{In fact, $X$ admits the $\bZ_5\times\bZ_5$ discrete symmetry. We will only consider one of them in this example.}
generated by 
\begin{eqnarray} \label{eqn:ac-ex2}
{\mathbb{Z}_5}:~~~~~~~g:z_i \to \alpha^{i} z_{i} 
\end{eqnarray} 
where $\alpha^5=1$. One can show that this action is fixed-point free on $X$, so the quotient manifold $\hat{X} =  X \Big/ \bZ_5$ is again a smooth CY threefold. The Euler characteristic of is $\chi(\hat{X}) = -40$ and the Hodge number is reduced to $h^{1,1}(\hat{X})=1$ and $h^{2,1}(\hat{X})=21$. 

\subsection*{Metric}

We will compute the RFM of $\hat{X}$ using Donaldson's algorithm. As discussed in section~\ref{sec:sec2}, the first step is to find an ample equivariant line bundle $\cL$ on $X$, such that the ansatz of K\"ahler potential defined in (\ref{eqn:ansatz1}) can be expanded using its invariant sections. 
In this example, we can simply take $\cL=\cO_X(1)$ which has an $\bZ_5$-equivariant structure given by %
\begin{equation} \label{eqn:phiex11}
\Phi_g = \left(
\begin{array}
[c]{ccccc}%
1 & 0&0& 0&0\\
0 & \alpha&0& 0&0\\
0 & 0&\alpha^2& 0&0\\
0 & 0&0& \alpha^3&0\\
0 & 0&0& 0&\alpha^4
\end{array}
\right) \;.
\end{equation}
The twists of this line bundle $\cL^k=\cO_X(k)$ with $k>0$ are also  $\bZ_5$-equivariant.

Let $\bC\left[z_0,z_1,\ldots,z_4\right]_k$ be the vector space of homogeneous polynomial of degree $k$ on $\bP^4$.  
The global sections of $\cO_X(k)$ are given by
\begin{equation*}
    H^0(X,\cO(k)) = \frac{\bC\left[z_0,z_1,\ldots,z_4\right]_k}{\left<P\right>}
\end{equation*}
where we have quotient the ideal generated by the defining polynomial $P$ for $k\geq 5$. 
While only $\bZ_5$ invariant sections of them are the global sections of the line bundle on the quotient manifold $\hat{X}$. 
By the equation (\ref{eqn:invSection2}) and the $\bZ_5$ action $\Phi_g$ given in (\ref{eqn:phiex11}), we obtain the invariant sections of $\cO_X(k)$. 
The number of these invariant sections $n_k$ for first a few $k$ are listed in the first row of Table~\ref{tab:ex1}. 
\begin{table}[!htp] \centering
  \renewcommand{\arraystretch}{1.3}
  \begin{tabular}{c|cccccccccc}
    $k$ &
    2 & 3 & 4 & 5 & 6 & 7 & 8&9&10&$\cdots$
    \\ \hline \strut
    $n_k$ &
    3& 7& 14& 25& 41 & 63& 92&129&175&$\cdots$
    \\
    $N_k$ &
    4& 12& 27& 55& 97 & 157& 238&343&475&$\cdots$
  \end{tabular}
  \caption{The number of invariant sections for the twisting line bundle $\cL=\cO_X(k)$ on the first row and for the twisted bundle 
  $\tilde{V}=V\otimes\cO_X(k)$ on the second row. }
  \label{tab:ex1}
\end{table}

As a cross check, the number of independent invariant sections $n_k$ can be determined from the index of $\cO_X(k)$. 
From the formula of line bundle cohomology on the quintic \cite{Constantin:2018hvl}, for $k>1$, the only non-vanishing cohomology is $H^0(X,\cO(k))$. 
Thus, by the equation (\ref{eqn:indexG}), we have
\begin{equation} \label{eqn:invSL-ex1}
    h^0(\hat{X},\cO(k))= \frac{1}{6}\, k^3 + \frac{5}{6} k+\delta_{k,0}, \quad k \geq 0\;.
\end{equation}
which gives the same numbers as $n_k$ listed in the Table~\ref{tab:ex1}.

\begin{figure}[!ht] \centering
\subfigure{\includegraphics[width=0.428\textwidth]{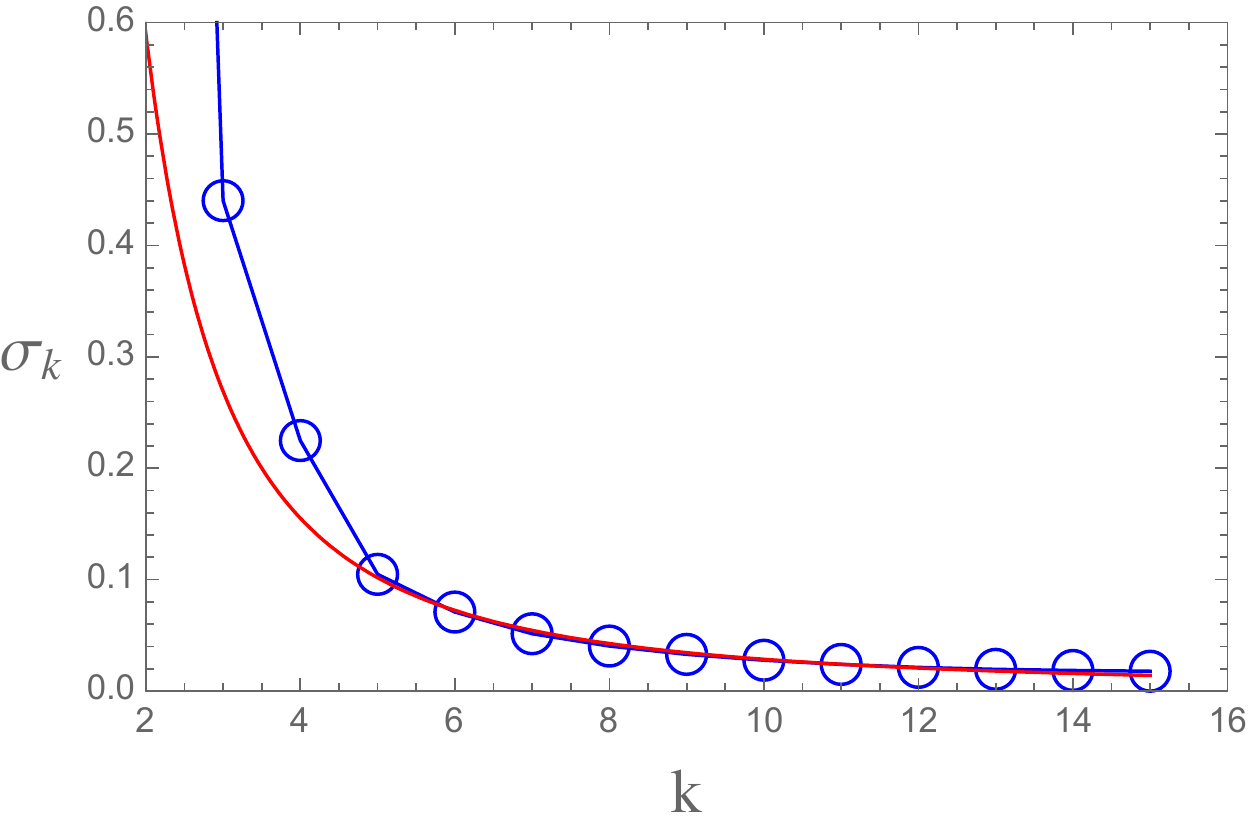}}
\subfigure{\includegraphics[width=0.428\textwidth]{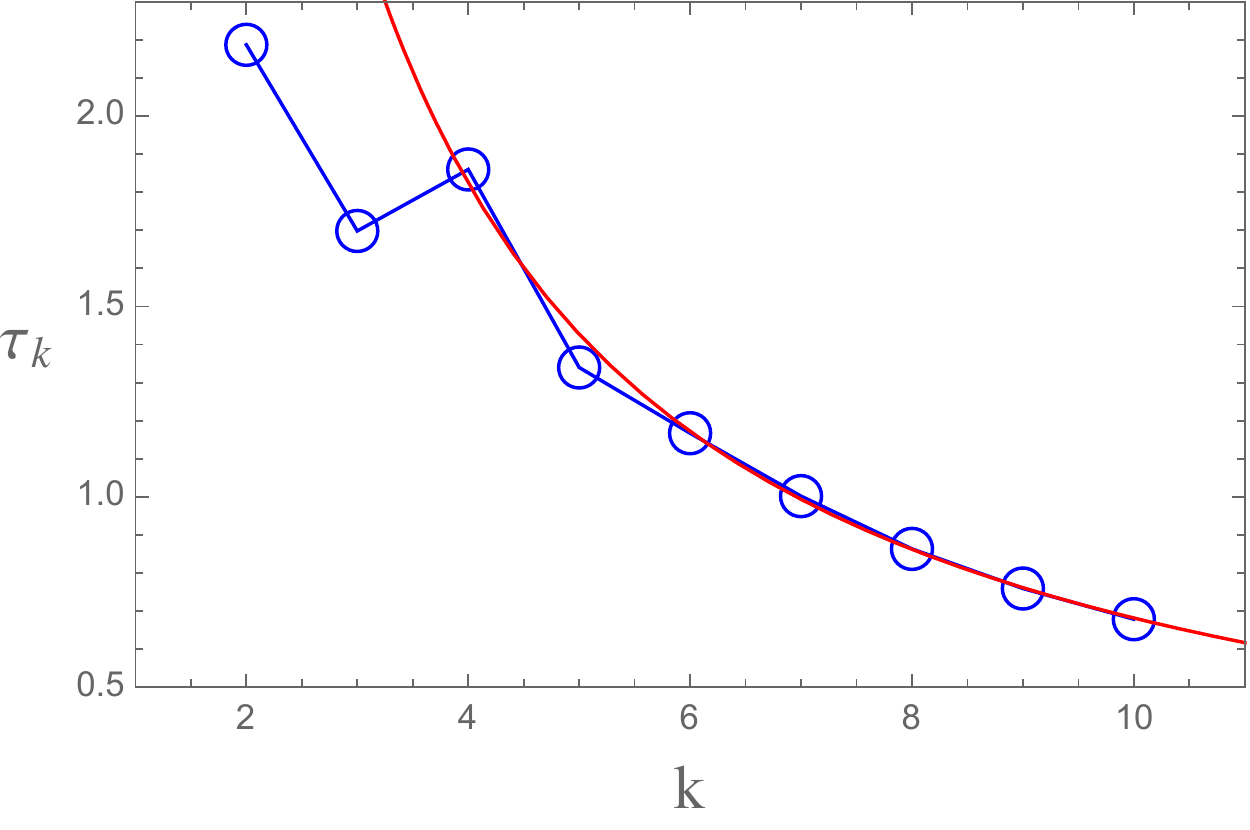}}
\caption{
The error measure of the balanced metric $\sigma_k$ from equation (\ref{eqn:error1}) and the balanced fiber metric $\tau_k$ from equation (\ref{eqn:error2}) versus the number of twists $k$ on the $\bZ_5$ quotient of the Quintic with the $SU(3)$ monad bundle, which scales as $\sigma_k \approx 2.24 k^{-2}+0.06 k^{-1}$ and $\tau_k \approx 3.16 k^{-2}+6.50 k^{-1}$ respectively. }  
\label{fig:ex2}
\end{figure}

With these invariant sections, we can get the ansatz for K\"ahler potential by (\ref{eqn:ansatz1}). Then, we perform the T-operator iteration starting from an arbitrary matrix $h_0$. 
With the method discussed in subsection~\ref{sec2:3}, the integral is evaluated numerically on a sampling of $2,500,000$ points. 
The balanced metric is computed in $60$ iterations for $k=2,3,\ldots, 15$. 
On the same sample of points, we evaluate the error function $\sigma_k$ using these balanced metric and plot the results in the left of Figure~\ref{fig:ex2}. 
As we can see that the error measure decreasing to zero in $k^{-2}$ that is consistent with the theoretical analysis in \cite{donaldson_2009,Douglas:2006rr}.
Also, it implies that the balanced metric will converge to the RFM on $\hat{X}$ as $k\to \infty$. 
We will take the balanced metric for $k=10$ as an approximation of RMF and use it to compute the error function for the fiber metric.

\subsection*{Connection}

Let's consider an $SU(3)$ holomorphic vector bundle $V$ on the quintic $X$ given by the following monad  
\begin{equation}\label{eqn:monadex1} 
0 
\rightarrow 
\cO_X(-4)
\rightarrow 
\cO_X(-1)^{\oplus 4} 
\stackrel{f}{\longrightarrow} 
V 
\rightarrow 
0 \;,
\end{equation} 
where $f \in H^0(X,\cO(3)^{\oplus 4})$ is the bundle map. 
Note that, in this section, we use the dual of the monad bundle defined in the equation (\ref{eqn:monad}). 
As we will see that the bundle $V$ can have a $\bZ_5$ equivariant structure and descend to an $SU(3)$ bundle $\hat{V}$ on the quotient manifold $\hat{X}$.  
We will use Donaldson's algorithm to compute the Hermitian-Einstein fiber metric of $\hat{V}$ so that the corresponding Chern connection gives the HYM connection.

We can show that the monad bundle $V$ can be $\bZ_5$ equivariant on $X$ for the following two reason. First, as we have seen that $\cO_X(1)$ is $\bZ_5$ equivariant and so are $B=\cO_X(-1)^{\oplus 4}$ and $C=\cO_X(-4)$.
Second, we take the bundle map $f$ in the monad above to be 
\begin{equation}\label{gen_map}
f^{T}=\left(
\begin{array}{l}
-z_0^3+4 z_2 z_3 z_0-3 z_1 z_4 z_0-5 z_1 z_2^2-3 z_2 z_4^2-3 z_1^2 z_3-2 z_3^2 z_4 \\
9 z_0^3+6 z_2 z_3 z_0+8 z_1 z_4 z_0-2 z_1 z_2^2+10 z_2 z_4^2+3 z_1^2 z_3-z_3^2 z_4 \\
z_0^3-4 z_2 z_3 z_0-3 z_1 z_4 z_0+4 z_2 z_4^2+4 z_1^2 z_3\\
-2 z_0^3+10 z_2 z_3 z_0+3 z_1 z_4
   z_0+6 z_1 z_2^2+z_2 z_4^2-2 z_1^2 z_3+5 z_3^2 z_4
\end{array}
\right)
\end{equation}
which is an invariant section of $H^0(X,\cO(3)^{\oplus 4})$ under the $\bZ_5$ symmetry. 
Thus, the bundle $V$ defined in (\ref{eqn:monadex1}) is a $\bZ_5$-equivariant bundle.

In fact, $V$ is a stable bundle because its dual bundle is defined by a positive monad \cite{Anderson:2007nc,Anderson:2008uw,Anderson:2008ex} that is stable by the Hoppe's criteria \cite{Hoppe1984}. 
 The corresponding bundle $\hat{V}$ on the quotient manifold is also stable and there exist an unique HYM connection. We will compute it using the Donaldson's algorithm.

To begin with, we use the equivariant line bundle $L=\cO(k)$ to twist the bundle $V$ and the twisted bundle $\tilde{V}=V \otimes \cO(k)$ is also equivariant given by the sequence below
\begin{equation}\label{eqn:monadex1-tw} 
0 
\rightarrow 
\cO_X(k-4)
\rightarrow 
\cO_X(k-1)^{\oplus 4} 
\stackrel{f}{\rightarrow} 
\tilde{V}
\rightarrow 
0 \;,
\end{equation} 
One can construct the global sections of $\tilde{V}$ by 
\begin{equation} \label{eqn:H0L-ex1}
    H^0(X,\tilde{V}) = \frac{H^0(X,\cO_X(k-1)^{\oplus 4} )}{f\left(H^0(X,\cO_X(k-4))\right)}
\end{equation}
While only the $\bZ_5$-invariant sections are the global sections of the bundle on $\hat{X}$, which can be obtained from the equation (\ref{eqn:invSection2}). Note that the section-wise maps of $V\otimes \cO_X(k)$ are defined by the equivariant structure of $V$ and $\cO_X(k)$. The number of these invariant sections $N_k$ for first a few $k$ are listed in the second row of Table~\ref{tab:ex1}.

We have worked out the number of independent invariant sections of an equivariant line bundle in (\ref{eqn:invSL-ex1}). 
With it, we can calculate the number of independent invariant sections for $V\otimes \cO_X(k)$
according to the equation (\ref{eqn:H0L-ex1}), which is given by 
\begin{equation}
\label{eqn:invSVL-ex1}
h^0(\hat{X},\hat{V}\otimes \hat{\cL^k})    = 
	\begin{cases}
	 \frac{2}{3} (k-1) \left(k^2-2 k+6\right)~, & 0< k < 4 \\[8pt] 
    \frac{1}{2} \left(k^3-7 k+20\right)-\delta_{k,4}~, & k\geq 4
	\end{cases}
\end{equation}
This is consistent with the result of $N_k$ listed in the Table~\ref{tab:ex1}.

With these invariant sections, we can write an ansatz for the fiber metric according to equation (\ref{eqn:ansatz2}). 
The balanced fiber metric will be computed using the T-operator iteration starting from an randomly chosen matrix $H^0_k$. 
The integration over $X$ is performed on a sample containing $400,000$ points.
After iterating $30$ times, the T-operator converges. In this way, we have computed the balanced fiber metric for $k=2, 3, \ldots, 10$.

To evaluate the accuracy, we compute the error function for each fiber metric on a smaller sample with $50,000$ points and the result is plotted in the right of Figure~\ref{fig:ex2}. 
The RFM used here is the balanced metric for $k=10$ computed in previous subsection. 
As one can easily see that the error function  converges to zero in $ \tau_k \sim k^{-2}$, which is expected in \cite{Douglas:2006hz}. 
The balanced fiber metric provide an good approximation of the Hermitian-Einstein fiber metric.

\subsection{An $SU(5)$ monad bundle on a $Z_3 \times Z_3$ quotient of Bi-cubic}

Let $X$ be a Bi-cubic Calabi-Yau threefold, 
\begin{equation*}
 X= \left[\begin{array}[c]{c}\mathbb{P}^2\\\mathbb{P}^2\end{array}\left|\begin{array}[c]{ccc}3 \\3
 \end{array}\right.  \right]^{2,83}\; ,
\end{equation*} 
defined as a hypersurface in $\mathbb{P}^2 \times \mathbb{P}^2$ by a generic degree $(3,3)$ homogeneous polynomial given by
\begin{align*}
    P&=-x_0^3y_0^3+x_0^3 y_1^3-x_0^3 y_2^3+x_0^3 y_0 y_1 y_2+x_2 x_0^2 y_0 y_1^2+x_1 x_0^2 y_0 y_2^2-x_2 x_0^2 y_1 y_2^2+x_1^2 x_0 y_1 y_2^2-x_1^2 x_0 y_0^2 y_2 -x_1^3 y_1^3\\
    &\quad+x_2^2 x_0 y_1^2 y_2 
    +x_1 x_2 x_0 y_0 y_1 y_2-x_1^3 y_0^3+x_2^3 y_0^3-x_2^3 y_1^3+x_1^3
   y_2^3-x_2^3 y_2^3-x_1 x_2^2 y_0 y_1^2+x_1^2 x_2 y_0^2 y_1 
   +x_1 x_2^2 y_0^2 y_2 \\
   & \quad+x_1^3 y_0 y_1 y_2+x_2^3 y_0 y_1 y_2+2 \left(x_0 x_2^2
   y_1 y_0^2+x_1^2 x_2 y_2^2 y_0+x_0^2 x_1 y_1^2 y_2\right)+3 \left(x_0^2 x_2 y_2 y_0^2+x_0 x_1^2 y_1^2 y_0+x_1 x_2^2 y_1 y_2^2\right) \;,
\end{align*}
where $\{x_i,y_i\}$ with $i=0,1,2$ are the homogeneous coordinates on $\mathbb{P}^2\times \mathbb{P}^{2}$. Note that this defining polynomial $P$ is chosen such that the $X$ admits a $\Gamma=\mathbb{Z}_3 \times \mathbb{Z}_3$ discrete symmetry, which are generated by
\begin{equation}\label{eqn:z3z3}
\begin{array}{llll}
\mathbb{Z}_3^{(1)}&:&x_k \to x_{k+1},&y_k \to y_{k+1}\\
\mathbb{Z}_3^{(2)}&:&x_k \to \alpha^k x_{k},&y_k \to \alpha^{-k} y_{k}~.
\end{array}
\end{equation}
where $\alpha=\exp (2\pi i/3)$. As discussed in \cite{Candelas:2007ac}, the action of $\Gamma$ is fix-point free and the quotient manifold $\hat{X}=X
\Big/ \Gamma
$ has Hodge number $h^{1,1}(\hat{X})=2$ and $h^{2,1}(\hat{X})=11$.

\subsection*{Metric}

We will compute the RFM on $\hat{X}$ with Donaldson's algorithm. 
To begin with, we will choose an equivariant ample line bundle on $X$. 
It is straightforward to see that 
any line bundle on $X$ can be generated by $L_1=\cO_X(1,0)$ and $L_2=\cO_X(0,1)$. However, as we will see that both of them are not $\Gamma$-equivariant.

Let's consider the first line bundle $L_1=O_X(1,0)$. 
A basis of $H^0(X,L_1)$ can be chosen to be $\{x_0,x_1,x_2\}$. From the equation (\ref{eqn:z3z3}), we can find the section-wise maps for both ${\mathbb{Z}_{3}}^{(1)}$ and ${\mathbb{Z}_{3}}^{(2)}$, which is given by
\begin{equation} \label{eqn:L1L2}
\Phi_{L_1,g_{1}}=\left(
\begin{array}
[c]{ccc}%
0 & 1&0\\
0&0 & 1\\
1&0 &0 
\end{array}
\right) \;,
\qquad 
\Phi_{L_1,g_2}=\left(
\begin{array}
[c]{ccc}%
1 & 0&0 \\
0 & \alpha^2&0\\
0&0& \alpha 
\end{array}
\right) \;.
\end{equation}
As we can check that the section-wise maps can define the $\bZ_3$-equivariant structure on $L_1$ independently. But they cannot be used to define a $\bZ_3 \times \bZ_3$ equivariant structure
because $\Phi_{L_1,g_1}$ and $\Phi_{L_1,g_2}$ do not commute \footnote{In fact, they form a representation of the order $27$ Heisenberg group, $H_{27}=(\mathbb{Z}_3 \times \mathbb{Z}_3) \rtimes \mathbb{Z}_3$ \cite{Anderson:2009mh}.}. 
Similarly, one can show that $L_2=\cO_X(0,1)$ is also not $\bZ_3 \times \bZ_3$ equivariant for the same reason. 


In fact, the simplest $\bZ_3 \times \bZ_3$ equivariant line bundle on $X$ is $\cL=\cO_X(1,1)$. Choose a basis of $H^0(X,\cO(1,1))$ to be $\{x_i y_j\}$ with $i,j=0,1,2$. Using the group action (\ref{eqn:z3z3}), we can find the section-wise maps $\Phi_{\cL,g_1}$ and $\Phi_{\cL,g_2}$,
\begin{equation} \label{eqn:phiL-ex2}
\Phi_{\cL,g_1}=
\left(
\begin{array}{ccccccccc}
 0 & 0 & 0 & 0 & 1 & 0 & 0 & 0 & 0 \\
 0 & 0 & 0 & 0 & 0 & 1 & 0 & 0 & 0 \\
 0 & 0 & 0 & 1 & 0 & 0 & 0 & 0 & 0 \\
 0 & 0 & 0 & 0 & 0 & 0 & 0 & 1 & 0 \\
 0 & 0 & 0 & 0 & 0 & 0 & 0 & 0 & 1 \\
 0 & 0 & 0 & 0 & 0 & 0 & 1 & 0 & 0 \\
 0 & 1 & 0 & 0 & 0 & 0 & 0 & 0 & 0 \\
 0 & 0 & 1 & 0 & 0 & 0 & 0 & 0 & 0 \\
 1 & 0 & 0 & 0 & 0 & 0 & 0 & 0 & 0 \\
\end{array}
\right)\;, \qquad 
\Phi_{\cL,g_2}=
\left(
\begin{array}{ccccccccc}
 1 & 0 & 0 & 0 & 0 & 0 & 0 & 0 & 0 \\
 0 & \alpha^2 & 0 & 0 & 0 & 0 & 0 & 0 & 0 \\
 0 & 0 & \alpha & 0 & 0 & 0 & 0 & 0 & 0 \\
 0 & 0 & 0 & \alpha & 0 & 0 & 0 & 0 & 0 \\
 0 & 0 & 0 & 0 & 1 & 0 & 0 & 0 & 0 \\
 0 & 0 & 0 & 0 & 0 & \alpha^2 & 0 & 0 & 0 \\
 0 & 0 & 0 & 0 & 0 & 0 & \alpha^2 & 0 & 0 \\
 0 & 0 & 0 & 0 & 0 & 0 & 0 & \alpha & 0 \\
 0 & 0 & 0 & 0 & 0 & 0 & 0 & 0 & 1 \\
\end{array}
\right)
\end{equation}
which commute with each other. Thus, $\cO_X(1,1)$ has a $\bZ_3 \times \bZ_3$ equivariant structure, 
which will descend to a line bundle $\hat{\cL}$ on the quotient manifold. We will use the invariant sections of $\cL^k=\cO_X(k,k)$ with $k\geq 1$ to define the K\"ahler potential in (\ref{eqn:ansatz1}) on $\hat{X}$.

Let the $\bC\left[x_0,x_1,x_2\right]_{k}$ be the vector space of homogeneous polynomial of degree $k$ on $\bP^2$.
The global sections of $\cO_X(k,k)$ are given by
\begin{equation*}
    H^0(X,\cO(k,k)) = \frac{\bC\left[x_0,x_1,x_2\right]_{k} \otimes \bC\left[y_0,y_1,y_2\right]_{k}}{\left<P\right>}
\end{equation*}
where we have quotient the ideal generated by the defining polynomial $P$ for $k\geq 3$. 
With the section-wise maps $\Phi_{\cL^k,g_1}$ and $\Phi_{\cL^k,g_2}$ induced from (\ref{eqn:phiL-ex2}), the invariant sections of $H^0(X,\cO(k,k))$ can be obtained from the equation (\ref{eqn:invSection2}). 
The number of invariant sections $n_k$ for the first a few twists is listed in the first row of Table~\ref{tab:ex2}. 
\begin{table}[!htp] \centering
  \renewcommand{\arraystretch}{1.3}
  \begin{tabular}{c|cccccccc}
    $k$ &
    2 & 3 & 4 & 5 & 6 & 7 & 8&$\cdots$
    \\ \hline \strut
    $n_k$ &
    4& 11& 24& 45& 76 & 119& 176&$\cdots$
    \\
    $N_k$ &
    12& 39& 97& 194& 341 & 548& 825&$\cdots$
  \end{tabular}
  \caption{The number of invariant sections for the twisting line bundle $L=\cO_X(k,k)$ (first row) and for the twisted bundle 
  $\tilde{V}=V\otimes\cO_X(k,k)$ (second row). }
  \label{tab:ex2}
\end{table}

The number of independent invariant sections can also be derived from the index of $\cO_X(k,k)$. The cohomology of a general line bundle $\cO(k_1,k_2)$ on $X$ has been studied in \cite{Constantin:2018hvl}. In the region of $k_1=k_2>0$ and $k_1=k_2<0$, the only non-trivial cohomology is $H^0(X,\cO(k_1,k_2))$. Thus, by the equation (\ref{eqn:indexG}), we have 
\begin{equation} \label{eqn:invSL-ex2}
  h^0(\hat{X},\hat{\cL^k}) = \frac{1}{3} k \left(k^2+2\right), \quad k\geq1 \;,
\end{equation}
which gives the same numbers as $n_k$ listed in the Table~\ref{tab:ex2}.

\begin{figure}[!ht] 
\centering
\subfigure{\includegraphics[width=0.4\textwidth]{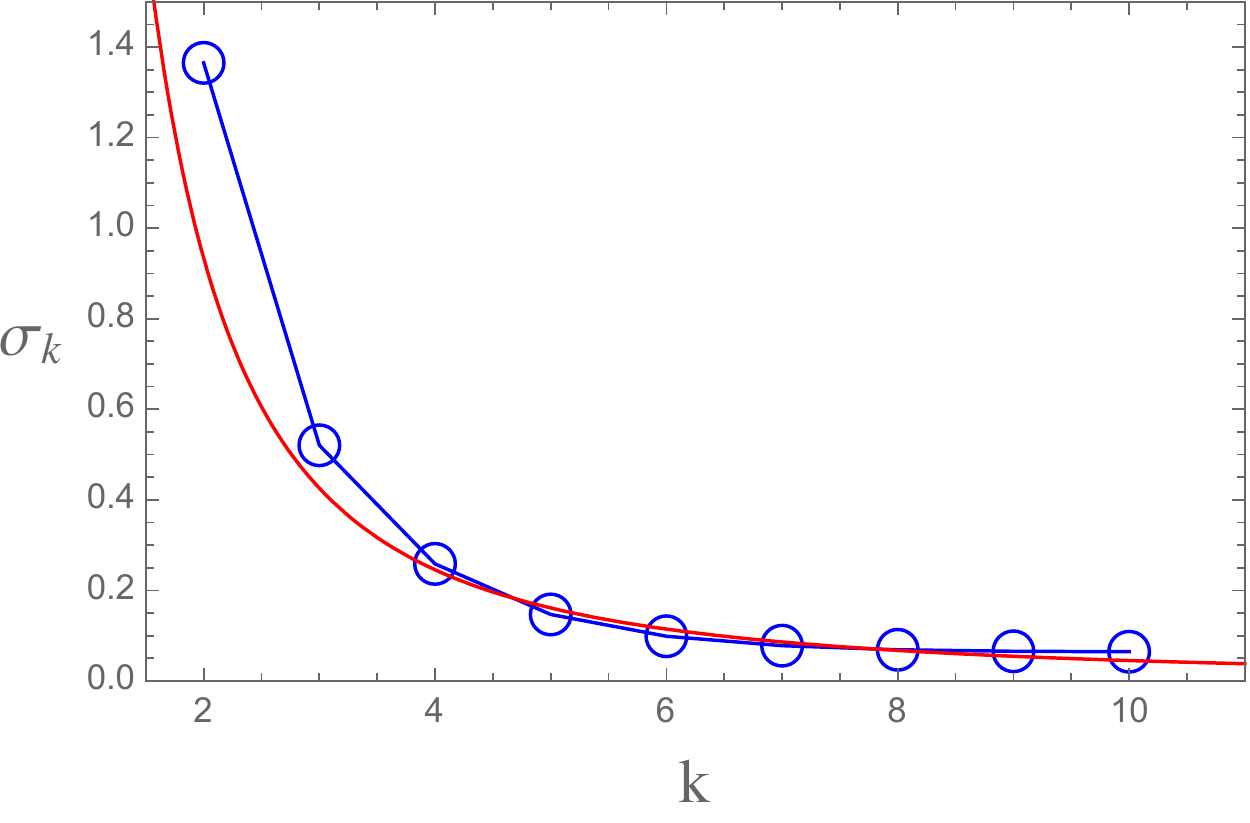}}
\subfigure{\includegraphics[width=0.4\textwidth]{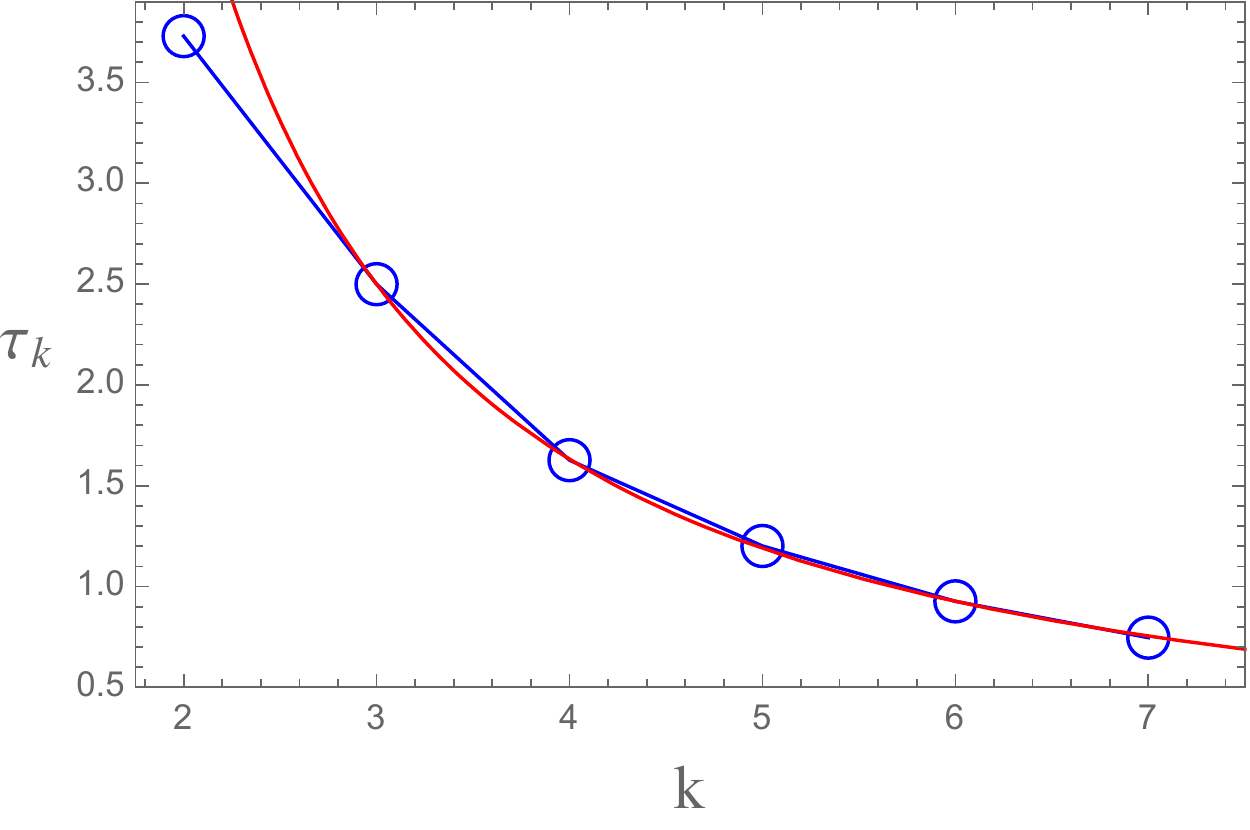}}
\caption{
The error measure of the balanced metric $\sigma_k$ from equation (\ref{eqn:error1}) and the balanced fiber metric $\tau_k$ from equation (\ref{eqn:error2}) versus the number of twists $k$ on the $\bZ_3 \times \bZ_3$ quotient of the Bi-cubic with the $SU(5)$ monad bundle, which scales as $\sigma_k \approx 3.54 k^{-2}+0.10 k^{-1}$ and $\tau_k \approx 11.64 k^{-2}+3.62 k^{-1}$ respectively.}
\label{fig:ex3}
\end{figure}

With these invariant sections, we can write down an ansatz of K\"ahler potential by the equation (\ref{eqn:ansatz1}). Then, we perform the T-operator iteration on a sampling containing $400,000$ points on $X$. The balanced metric is computed in $30$ iterations for $k=1,2,\ldots, 10$. 
We evaluate the accuracy of these balanced metric on a smaller sample with $50,000$ points and plot the results in the left of the Figure~\ref{fig:ex3}. 
The error function decreases with the increase of $k$ in a manner of $\sigma_k \sim k^{-2}$, which is again expected from the analysis made in \cite{donaldson_2009,Douglas:2006rr}.  
We will take the balanced metric for $k=8$ as an approximation of RMF and use it to compute the error function for the fiber metric.

\subsection*{Connection}

In this subsection, we will consider an $SU(5)$ vector bundle which also admits a $\mathbb{Z}_3 \times \mathbb{Z}_3$ equivariant structure and, hence, descends to a bundle $\hat{V}$ on $\hat{X}$. We will use Donaldson's algorithm to compute the HYM connection of $\hat{V}$. Consider an $SU(5)$ holomorphic vector bundle constructed by the monad below
\cite{Anderson:2009mh}
\begin{equation}\label{eqn:monadex2}  
 0 \to  \cO_X(-3,-3) 
\to \cO_X(-1,0)^{\oplus 3} \oplus \cO_X(0,-1)^{\oplus 3} \stackrel{f}{\rightarrow} 
V\to 0 \;,
\end{equation}
where we denote 
\begin{equation*}
B=\cO_X(-1,0)^{\oplus 3} \oplus \cO_X(0,-1)^{\oplus 3}\;, \qquad C= \cO_X(-3,-3)\;.
\end{equation*}
Here $f \in H^0(X,B^* \otimes C)$ is a bundle map. For convenience, we write it as $f=f_1 \oplus f_2$ with 
\begin{equation} \label{eqn:f1f2}
    f_1 \in H^0(X, \cO(2,3)^{\oplus 3} )\;, \qquad 
    f_2\in H^0(X,\cO(3,2)^{\oplus 3}) \;.
\end{equation}
This shows that $f_1$ and $f_2$ are vectors containing 3 homogeneous polynomials of degree $(2,3)$ and $(3,2)$ respectively. 
As we will explain later, they will be chosen such that $V$ has the $\bZ_3 \times \bZ_3$ equivariant structure.

The $\bZ_3 \times \bZ_3$ equivariant structure of $V$ is defined as follows.
The line bundle $\cO_X(1,1)$ is equivariant and therefore $C$ admits an equivariant structure. 
While for $B$, we have seen that the individual line bundle $ \cO_X(0,-1)$ and $ \cO_X(-1,0)$ are not equivariant, but the direct sum of them could be equivariant.


For example, consider the direct sum of three line bundles $L_1^{\oplus 3}=\cO_X(1,0)^{\oplus 3}$. As shown in \cite{Anderson:2009nt}, it has a $\bZ_3 \times \bZ_3$ equivariant structure with the following non-trivial bundle maps 
\begin{equation} \label{eqn:bmorphismZ3Z3}
\phi_{L_1^{\oplus 3},g_1} =
\left(
\begin{array}
[c]{ccc}%
0 & 1&0\\
0&0 & 1\\
1&0 &0 
\end{array}
\right), \qquad 
\phi_{L_1^{\oplus 3},g_2}
\left(
\begin{array}
[c]{ccc}%
1 & 0&0 \\
0 & \alpha&0\\
0&0& \alpha^2 
\end{array}
\right) \;.
\end{equation}
With these bundle maps, the section-wise maps are given by \cite{Anderson:2009mh}
\begin{equation}
    \Phi_{L_1^{\oplus 3},g_{1}}(s) = \phi_{L_1^{\oplus 3},g_1} \otimes \Phi_{L_1,g_1}\;, \qquad 
\Phi_{L_1^{\oplus 3},g_{2}}(s)  = \phi_{L_1^{\oplus 3},g_2} \otimes \Phi_{L_1,g_2}\;
\end{equation}
where $\Phi_{L_1,g_1}$ and $\Phi_{L_1,g_2}$ are defined in (\ref{eqn:L1L2}). 
As one can explicitly check that they commute with each other and form a representation of $\bZ_3 \times \bZ_3$.
Similarly, for $L_2^{\oplus 3}=\cO_X(0,1)^{\oplus 3}$, one can construct actions $\Phi_{L_2^{\oplus 3},g_{1}}$ and $\Phi_{L_2^{\oplus 3},g_{2}}$ with the same bundle morphism given in (\ref{eqn:bmorphismZ3Z3}), which forms a representation of $\bZ_3 \times \bZ_3$ and defines an equivariant structure on $\cL_2^{\oplus 3}$. In this way, we have shown that both $B$ and $C$ are $\bZ_3 \times \bZ_3$ equivariant in the monad (\ref{eqn:monadex2}).


In order for $V$ to be equivariant, the bundle map $f_1$ and $f_2$ in (\ref{eqn:f1f2}) should be an invariant section in $H^0(X,\cO(2,3)^{\oplus 3})$ and $H^0(X,\cO(3,2)^{\oplus 3})$ respectively. Let's focus on how to choose the first map $f_1$. The global section of $\cO(2,3)^{\oplus 3}$ is given by 
\begin{equation*}
    H^0(X,\cO(2,3)^{\oplus 3}) = \bC\left[x_0,x_1,x_2\right]_{2} \otimes \bC\left[y_0,y_1,y_2\right]_{3}\;,
\end{equation*}
With the section-wise maps induced from $\Phi_{L_1^{\oplus 3},g_i}$ and $\Phi_{C,g_i}$, we calculate the invariant sections by equation (\ref{eqn:invSection2}). 
We choose the map $f_1=(f_{11},f_{12},f_{13})$ to be a generic invariant section with components 
\begin{equation}\label{eqn:fcom-ex2}
\begin{aligned} 
f_{11} = &   a_{20} x_0^2 y_0^3+a_{16} x_1 x_2 y_0^3+a_5 x_2^2 y_1 y_0^2+a_6 x_0 x_1 y_1 y_0^2+a_9 x_1^2 y_2 y_0^2+a_{13} x_0 x_2 y_2 y_0^2+a_{14}
  x_1^2 y_1^2 y_0  \\
&  +a_8 x_2^2 y_2^2 y_0 
   +a_4 x_0 x_1 y_2^2 y_0+a_2 x_0^2 y_1 y_2 y_0
  +a_1 x_1 x_2 y_1 y_2 y_0+a_{17}
  x_0^2 y_1^3+a_{18} x_1 x_2 y_1^3+a_{15} x_0^2 y_2^3  \\
  &+a_{19} x_1 x_2 y_2^3+a_{11} x_1^2 y_1 y_2^2+a_{12} x_0 x_2 y_1 y_2^2+a_3 x_2^2
  y_1^2 y_2+a_7 x_0 x_1 y_1^2 y_2 +a_{10} x_0 x_2 y_1^2 y_0\\ 
  f_{12} =&a_{15} x_1^2 y_0^3+a_{19} x_0 x_2 y_0^3+a_8 x_0^2 y_1 y_0^2+a_4 x_1 x_2 y_1 y_0^2+a_{11} x_2^2 y_2 y_0^2+a_{12} x_0 x_1 y_2 y_0^2+a_9 x_2^2 y_1^2 y_0  \\
 &+a_3 x_0^2 y_2^2 y_0+a_7 x_1 x_2 y_2^2 y_0+a_2 x_1^2 y_1 y_2 y_0+a_1 x_0 x_2 y_1 y_2 y_0+a_{20} x_1^2 y_1^3+a_{16} x_0 x_2 y_1^3+a_{17} x_1^2 y_2^3 \\
 &+a_{14} x_2^2 y_1 y_2^2+a_{10} x_0 x_1 y_1 y_2^2+a_5 x_0^2
  y_1^2 y_2+a_6 x_1 x_2 y_1^2 y_2 +a_{13} x_0 x_1 y_1^2 y_0 +a_{18} x_0 x_2 y_2^3 \\ 
  f_{13} =&   a_{17} x_2^2 y_0^3+a_{18} x_0 x_1 y_0^3+a_3 x_1^2 y_1 y_0^2+a_7 x_0 x_2 y_1 y_0^2+a_{14} x_0^2 y_2 y_0^2+a_{10} x_1 x_2 y_2 y_0^2+a_{11}
  x_0^2 y_1^2 y_0 \\
  &+a_5 x_1^2 y_2^2 y_0+a_6 x_0 x_2 y_2^2 y_0+a_2 x_2^2 y_1 y_2 y_0+a_1 x_0 x_1 y_1 y_2 y_0+a_{15}
  x_2^2 y_1^3+a_{19} x_0 x_1 y_1^3+a_{20} x_2^2 y_2^3 \\
 & +a_9 x_0^2 y_1 y_2^2+a_{13} x_1 x_2 y_1 y_2^2+a_8 x_1^2 y_1^2
  y_2+a_4 x_0 x_2 y_1^2 y_2+a_{12} x_1 x_2 y_1^2 y_0+a_{16} x_0 x_1 y_2^3
\end{aligned}
\end{equation}
where $a_i$, $i=1,2,\ldots 20$ are the arbitrary coefficients. 
Similarly, for $\cO(3,2)^{\oplus 3}$, we can also construct a section-wise map and work out the invariant sections. 
The map $f_2=(f_{21},f_{22},f_{23})$ is chosen to be a generic such invariant section with components given by the equation (\ref{eqn:fcom-ex2}) with homogeneous coordinates $x_i$ and $y_i$ are switched. 
Thus, we have shown that the monad bundle $V$ is $\bZ_3 \times \bZ_3$ equivariant and will descend to a bundle $\hat{V}$ on $\hat{X}$. We will use the Donaldson's algorithm to solve its HYM connection.

Let's consider the twist of the bundle $V$ by an equivariant line bundle $L=\cO_X(k,k)$ with $k \geq 1$. 
The twisted bundle $\tilde{V}=V \otimes \cO_X(k,k)$ is obviously equivariant given by 
\begin{equation}\label{eqn:monadex2-tw} 
0 \rightarrow \cO_X(k-3,k-3) \rightarrow \cO_X(k-1,k)^{\oplus 3} \oplus \cO_X(k,k-1)^{\oplus 3} \stackrel{f}{\longrightarrow} \hat{V}  \rightarrow 0 \;. 
\end{equation} 
One can construct the global sections of $\tilde{V}$ by 
\begin{equation} \label{eqn:H0L-ex2}
    H^0(X,\tilde{V}) = \frac{H^0(X,\cO_X(k-1,k)^{\oplus 3} \oplus \cO_X(k,k-1)^{\oplus 3} )}{f\left(H^0(X,\cO_X(k-3,k-3))\right)}
\end{equation}
Among them, only the $\Gamma$-invariant sections are the global sections of the bundle on $\hat{X}$, which can be obtained from the equation (\ref{eqn:invSection2}). Note that the section-wise maps of $V\otimes \cO_X(k,k)$ are induced from the equivariant structure of the bundle $V$ and the line bundle in equation (\ref{eqn:phiL-ex2}). The number of these invariant sections $N_k$ for first a few $k$ are listed in the second row of Table~\ref{tab:ex2}.

We have worked out the number of independent invariant sections of an equivariant line bundle in (\ref{eqn:invSL-ex2}). 
With it, we can calculate the number of independent invariant sections for $V\otimes \cO_X(k,k)$
according to the equation (\ref{eqn:H0L-ex1}), which is given by 
\begin{equation}
\label{eqn:invSVL-ex2}
h^0(\hat{X},\hat{V}\otimes \hat{\cL^k})    = 
	\begin{cases}
	 \left(k^2-k+2\right)(2 k-1)~, & 0< k < 3 \\[8pt] 
	\frac{1}{3} \left ( 5 k^3 - 14k + 27\right)-\delta_{k,3}~,&  k \geq 3
	\end{cases}
\end{equation}
This is consistent with the result of $N_k$ listed in the Table~\ref{tab:ex2}.

According to the equation (\ref{eqn:ansatz2}), we write an ansatz for the fiber metric of $\hat{\tilde{V}}$ using these invariant sections. Then, perform the T-operator iteration on a sample of $50,000$ points. 
After $30$ times iteration, we obtain the balanced fiber metric for $k=2,3,\ldots,8$. 
To evaluate the accuracy, we compute the error function for each fiber metric on the same sample and the result is plotted in the right of Figure~\ref{fig:ex3}. 
The RFM used here is the balanced metric for $k=8$ computed in previous subsection. 
We find that, as the increase of $k$, the error function decreases as $ \tau_k \sim k^{-2}$ that is consistent with the analysis made in \cite{Douglas:2006hz}. 
Therefore the Donaldson's algorithm provides an effective and controllable way to compute the Hermitian-Einstein fiber metric of $\hat{V}$.

\section{Conclusion and Outlook}
\label{sec:sec4}

In this work, we extended the generalized Donaldson algorithm to holomorphic poly-stable bundles on quotient manifolds and presented a systematic way to approximate the HYM connection numerically. 
The key step in the algorithm is to define the equivariant structure of upstairs bundles and use it to find the global sections of the downstairs bundles. 
To illustrate our method, we considered an $SU(3)$ monad bundle on a $Z_5$ quotient of quintic and an $SU(5)$ monad bundle on a $Z_3 \times Z_3$ quotient of Bi-cubic. 
The equivariant structure of the former is defined with the trivial bundle morphism while the one for the later involves non-trivial bundle morphism. 
With the generalized Donaldson algorithm, we studied the HYM connection for both of examples. Our numerical results converged as expected and provide a good approximation of the HYM connections.

Given a holomorphic bundle, determining its stability is in general difficult. 
It is found in \cite{Anderson:2010ke,Anderson:2011ed} that the T-operator iteration behaves differently for poly-stable and unstable bundles and the generalized Donaldson's algorithm can be used to check the stability of the bundles on simply connected CY manifolds. 
It is interesting to see if this criterion is also true for bundles defined on the quotient manifolds. 
For CY manifold with $h^{1,1}>1$, the presence of the holomorphic bundle can introduce a substructure in K\"ahler cone, which separating the stable and unstable regions are called ``stability wall" \cite{Anderson:2009sw, Anderson:2009nt, Anderson:2010tc}. 
With our method, it is also interesting to study the stability wall caused by bundles on the quotient manifolds. 
Finally, realistic heterotic models are realized on the quotient CY manifold. With the HYM connections constructed in this paper, we will compute the matter-field K\"ahler potential and physical Yukawa couplings in the CY compactification.

\section*{Acknowledgments}

We would like to thank Juntao Wang for valuable discussions and careful reading of the manuscript. 
The work is supported by the fellowship of China Postdoctoral Science Foundation NO.2022M720507. 
The authors acknowledge Advanced Research Computing at Virginia Tech for providing computational resources and technical support that have contributed to the results reported within this paper. 
In addition, we would like to thank the Tsinghua Sanya International Mathematics Forum (TSIMF) for hospitality where part of this work was completed.

\bibliographystyle{unsrt}
\bibliography{main}

\begin{thebibliography}{10}

\bibitem{Candelas:1985en}
P.~Candelas, Gary~T. Horowitz, Andrew Strominger, and Edward Witten.
\newblock {Vacuum configurations for superstrings}.
\newblock {\em Nucl. Phys. B}, 258:46--74, 1985.

\bibitem{Yau1978OnTR}
Shing-Tung Yau.
\newblock On the ricci curvature of a compact kahler manifold and the complex
  monge-ampere equation, i*.
\newblock {\em Communications on Pure and Applied Mathematics}, 31:339--411,
  1978.

\bibitem{Green:1987mn}
Michael~B. Green, J.~H. Schwarz, and Edward Witten.
\newblock {\em {SUPERSTRING THEORY. VOL. 2: LOOP AMPLITUDES, ANOMALIES AND
  PHENOMENOLOGY}}.
\newblock 7 1988.

\bibitem{Donaldson:1985zz}
S.~K. Donaldson.
\newblock {ANTI SELF-DUAL YANG-MILLS CONNECTIONS OVER COMPLEX ALGEBRAIC
  SURFACES AND STABLE VECTOR BUNDLES}.
\newblock {\em Proc. Lond. Math. Soc.}, 50:1--26, 1985.

\bibitem{UYau}
K.~Uhlenbeck and S.~T. Yau.
\newblock On the existence of hermitian-yang-mills connections in stable vector
  bundles.
\newblock {\em Communications on Pure and Applied Mathematics},
  39(S1):S257--S293, 1986.

\bibitem{Anderson:2010mh}
Lara~B. Anderson, James Gray, Andre Lukas, and Burt Ovrut.
\newblock {Stabilizing the Complex Structure in Heterotic Calabi-Yau Vacua}.
\newblock {\em JHEP}, 02:088, 2011.

\bibitem{Anderson:2011ty}
Lara~B. Anderson, James Gray, Andre Lukas, and Burt Ovrut.
\newblock {The Atiyah Class and Complex Structure Stabilization in Heterotic
  Calabi-Yau Compactifications}.
\newblock {\em JHEP}, 10:032, 2011.

\bibitem{Anderson:2011cza}
Lara~B. Anderson, James Gray, Andre Lukas, and Burt Ovrut.
\newblock {Stabilizing All Geometric Moduli in Heterotic Calabi-Yau Vacua}.
\newblock {\em Phys. Rev. D}, 83:106011, 2011.

\bibitem{Anderson:2013qca}
Lara~B. Anderson, James Gray, Andre Lukas, and Burt Ovrut.
\newblock {Vacuum Varieties, Holomorphic Bundles and Complex Structure
  Stabilization in Heterotic Theories}.
\newblock {\em JHEP}, 07:017, 2013.

\bibitem{Anderson:2009sw}
Lara~B. Anderson, James Gray, Andre Lukas, and Burt Ovrut.
\newblock {The Edge Of Supersymmetry: Stability Walls in Heterotic Theory}.
\newblock {\em Phys. Lett. B}, 677:190--194, 2009.

\bibitem{Anderson:2009nt}
Lara~B. Anderson, James Gray, Andre Lukas, and Burt Ovrut.
\newblock {Stability Walls in Heterotic Theories}.
\newblock {\em JHEP}, 09:026, 2009.

\bibitem{Cui:2020ijv}
Wei Cui and Mohsen Karkheiran.
\newblock {Heterotic Complex Structure Moduli Stabilization for Elliptically
  Fibered Calabi-Yau Manifolds}.
\newblock {\em JHEP}, 03:281, 2021.

\bibitem{Greene:1986bm}
Brian~R. Greene, Kelley~H. Kirklin, Paul~J. Miron, and Graham~G. Ross.
\newblock {A Three Generation Superstring Model. 1. Compactification and
  Discrete Symmetries}.
\newblock {\em Nucl. Phys. B}, 278:667--693, 1986.

\bibitem{Greene:1986jb}
Brian~R. Greene, Kelley~H. Kirklin, Paul~J. Miron, and Graham~G. Ross.
\newblock {A Three Generation Superstring Model. 2. Symmetry Breaking and the
  Low-Energy Theory}.
\newblock {\em Nucl. Phys. B}, 292:606--652, 1987.

\bibitem{Bouchard:2005ag}
Vincent Bouchard and Ron Donagi.
\newblock {An SU(5) heterotic standard model}.
\newblock {\em Phys. Lett. B}, 633:783--791, 2006.

\bibitem{Braun:2005zv}
Volker Braun, Yang-Hui He, Burt~A. Ovrut, and Tony Pantev.
\newblock {Vector bundle extensions, sheaf cohomology, and the heterotic
  standard model}.
\newblock {\em Adv. Theor. Math. Phys.}, 10(4):525--589, 2006.

\bibitem{Braun:2005bw}
Volker Braun, Yang-Hui He, Burt~A. Ovrut, and Tony Pantev.
\newblock {A Standard model from the E(8) x E(8) heterotic superstring}.
\newblock {\em JHEP}, 06:039, 2005.

\bibitem{Braun:2005ux}
Volker Braun, Yang-Hui He, Burt~A. Ovrut, and Tony Pantev.
\newblock {A Heterotic standard model}.
\newblock {\em Phys. Lett. B}, 618:252--258, 2005.

\bibitem{Anderson:2009mh}
Lara~B. Anderson, James Gray, Yang-Hui He, and Andre Lukas.
\newblock {Exploring Positive Monad Bundles And A New Heterotic Standard
  Model}.
\newblock {\em JHEP}, 02:054, 2010.

\bibitem{Braun:2011ni}
Volker Braun, Philip Candelas, Rhys Davies, and Ron Donagi.
\newblock {The MSSM Spectrum from (0,2)-Deformations of the Heterotic Standard
  Embedding}.
\newblock {\em JHEP}, 05:127, 2012.

\bibitem{Anderson:2011ns}
Lara~B. Anderson, James Gray, Andre Lukas, and Eran Palti.
\newblock {Two Hundred Heterotic Standard Models on Smooth Calabi-Yau
  Threefolds}.
\newblock {\em Phys. Rev. D}, 84:106005, 2011.

\bibitem{Anderson:2012yf}
Lara~B. Anderson, James Gray, Andre Lukas, and Eran Palti.
\newblock {Heterotic Line Bundle Standard Models}.
\newblock {\em JHEP}, 06:113, 2012.

\bibitem{Anderson:2013xka}
Lara~B. Anderson, Andrei Constantin, James Gray, Andre Lukas, and Eran Palti.
\newblock {A Comprehensive Scan for Heterotic SU(5) GUT models}.
\newblock {\em JHEP}, 01:047, 2014.

\bibitem{Gray:2019tzn}
James Gray and Juntao Wang.
\newblock {Jumping Spectra and Vanishing Couplings in Heterotic Line Bundle
  Standard Models}.
\newblock {\em JHEP}, 11:073, 2019.

\bibitem{Strominger:1985ks}
Andrew Strominger.
\newblock {Yukawa Couplings in Superstring Compactification}.
\newblock {\em Phys. Rev. Lett.}, 55:2547, 1985.

\bibitem{Distler:1987ee}
Jacques Distler and Brian~R. Greene.
\newblock {Aspects of (2,0) String Compactifications}.
\newblock {\em Nucl. Phys. B}, 304:1--62, 1988.

\bibitem{Candelas:1987se}
P.~Candelas.
\newblock {Yukawa Couplings Between (2,1) Forms}.
\newblock {\em Nucl. Phys. B}, 298:458, 1988.

\bibitem{Candelas:1987rx}
Philip Candelas and Sunny Kalara.
\newblock {Yukawa Couplings for a Three Generation Superstring
  Compactification}.
\newblock {\em Nucl. Phys. B}, 298:357--368, 1988.

\bibitem{Distler:1995bc}
Jacques Distler and Shamit Kachru.
\newblock {Duality of (0,2) string vacua}.
\newblock {\em Nucl. Phys. B}, 442:64--74, 1995.

\bibitem{Braun:2006me}
Volker Braun, Yang-Hui He, and Burt~A. Ovrut.
\newblock {Yukawa couplings in heterotic standard models}.
\newblock {\em JHEP}, 04:019, 2006.

\bibitem{Bouchard:2006dn}
Vincent Bouchard, Mirjam Cvetic, and Ron Donagi.
\newblock {Tri-linear couplings in an heterotic minimal supersymmetric standard
  model}.
\newblock {\em Nucl. Phys. B}, 745:62--83, 2006.

\bibitem{Anderson:2009ge}
Lara~B. Anderson, James Gray, Dan Grayson, Yang-Hui He, and Andre Lukas.
\newblock {Yukawa Couplings in Heterotic Compactification}.
\newblock {\em Commun. Math. Phys.}, 297:95--127, 2010.

\bibitem{Anderson:2010tc}
Lara~B. Anderson, James Gray, and Burt Ovrut.
\newblock {Yukawa Textures From Heterotic Stability Walls}.
\newblock {\em JHEP}, 05:086, 2010.

\bibitem{Blesneag:2015pvz}
Stefan Blesneag, Evgeny~I. Buchbinder, Philip Candelas, and Andre Lukas.
\newblock {Holomorphic Yukawa Couplings in Heterotic String Theory}.
\newblock {\em JHEP}, 01:152, 2016.

\bibitem{Blesneag:2016yag}
Stefan Blesneag, Evgeny~I. Buchbinder, and Andre Lukas.
\newblock {Holomorphic Yukawa Couplings for Complete Intersection Calabi-Yau
  Manifolds}.
\newblock {\em JHEP}, 01:119, 2017.

\bibitem{Anderson:2021unr}
Lara~B. Anderson, James Gray, Magdalena Larfors, Matthew Magill, and Robin
  Schneider.
\newblock {Generalized Vanishing Theorems for Yukawa Couplings in Heterotic
  Compactifications}.
\newblock {\em JHEP}, 05:085, 2021.

\bibitem{Blesneag:2018ygh}
\c{S}tefan Blesneag, Evgeny~I. Buchbinder, Andrei Constantin, Andre Lukas, and
  Eran Palti.
\newblock {Matter field K\"ahler metric in heterotic string theory from
  localisation}.
\newblock {\em JHEP}, 04:139, 2018.

\bibitem{donaldson_2009}
S.~K. Donaldson.
\newblock Some numerical results in complex differential geometry.
\newblock {\em Pure and Applied Mathematics Quarterly}, 5(2):571--618, 2009.

\bibitem{Douglas:2006rr}
Michael~R. Douglas, Robert~L. Karp, Sergio Lukic, and Rene Reinbacher.
\newblock {Numerical Calabi-Yau metrics}.
\newblock {\em J. Math. Phys.}, 49:032302, 2008.

\bibitem{Braun:2007sn}
Volker Braun, Tamaz Brelidze, Michael~R. Douglas, and Burt~A. Ovrut.
\newblock {Calabi-Yau Metrics for Quotients and Complete Intersections}.
\newblock {\em JHEP}, 05:080, 2008.

\bibitem{Douglas:2006hz}
Michael~R. Douglas, Robert~L. Karp, Sergio Lukic, and Rene Reinbacher.
\newblock {Numerical solution to the hermitian Yang-Mills equation on the
  Fermat quintic}.
\newblock {\em JHEP}, 12:083, 2007.

\bibitem{Anderson:2010ke}
Lara~B. Anderson, Volker Braun, Robert~L. Karp, and Burt~A. Ovrut.
\newblock {Numerical Hermitian Yang-Mills Connections and Vector Bundle
  Stability in Heterotic Theories}.
\newblock {\em JHEP}, 06:107, 2010.

\bibitem{Anderson:2011ed}
Lara~B. Anderson, Volker Braun, and Burt~A. Ovrut.
\newblock {Numerical Hermitian Yang-Mills Connections and Kahler Cone
  Substructure}.
\newblock {\em JHEP}, 01:014, 2012.

\bibitem{Headrick:2009jz}
Matthew Headrick and Ali Nassar.
\newblock {Energy functionals for Calabi-Yau metrics}.
\newblock {\em Adv. Theor. Math. Phys.}, 17(5):867--902, 2013.

\bibitem{Headrick:2005ch}
Matthew Headrick and Toby Wiseman.
\newblock {Numerical Ricci-flat metrics on K3}.
\newblock {\em Class. Quant. Grav.}, 22:4931--4960, 2005.

\bibitem{Cui:2019uhy}
Wei Cui and James Gray.
\newblock {Numerical Metrics, Curvature Expansions and Calabi-Yau Manifolds}.
\newblock {\em JHEP}, 05:044, 2020.

\bibitem{Ashmore:2019wzb}
Anthony Ashmore, Yang-Hui He, and Burt~A. Ovrut.
\newblock {Machine Learning Calabi\textendash{}Yau Metrics}.
\newblock {\em Fortsch. Phys.}, 68(9):2000068, 2020.

\bibitem{Douglas:2020hpv}
Michael~R. Douglas, Subramanian Lakshminarasimhan, and Yidi Qi.
\newblock {Numerical Calabi-Yau metrics from holomorphic networks}.
\newblock 12 2020.

\bibitem{Anderson:2020hux}
Lara~B. Anderson, Mathis Gerdes, James Gray, Sven Krippendorf, Nikhil Raghuram,
  and Fabian Ruehle.
\newblock {Moduli-dependent Calabi-Yau and SU(3)-structure metrics from Machine
  Learning}.
\newblock {\em JHEP}, 05:013, 2021.

\bibitem{Jejjala:2020wcc}
Vishnu Jejjala, Damian~Kaloni Mayorga~Pena, and Challenger Mishra.
\newblock {Neural network approximations for Calabi-Yau metrics}.
\newblock {\em JHEP}, 08:105, 2022.

\bibitem{Ashmore:2021rlc}
Anthony Ashmore, Rehan Deen, Yang-Hui He, and Burt~A. Ovrut.
\newblock {Machine learning line bundle connections}.
\newblock {\em Phys. Lett. B}, 827:136972, 2022.

\bibitem{Larfors:2021pbb}
Magdalena Larfors, Andre Lukas, Fabian Ruehle, and Robin Schneider.
\newblock {Learning Size and Shape of Calabi-Yau Spaces}.
\newblock 11 2021.

\bibitem{Larfors:2022nep}
Magdalena Larfors, Andre Lukas, Fabian Ruehle, and Robin Schneider.
\newblock {Numerical metrics for complete intersection and
  Kreuzer\textendash{}Skarke Calabi\textendash{}Yau manifolds}.
\newblock {\em Mach. Learn. Sci. Tech.}, 3(3):035014, 2022.

\bibitem{Ashmore:2021ohf}
Anthony Ashmore, Lucille Calmon, Yang-Hui He, and Burt~A. Ovrut.
\newblock {Calabi-Yau Metrics, Energy Functionals and Machine-Learning}.
\newblock 12 2021.

\bibitem{Berglund:2022gvm}
Per Berglund, Giorgi Butbaia, Tristan H\"ubsch, Vishnu Jejjala, Dami\'an
  Mayorga Pe\~na, Challenger Mishra, and Justin Tan.
\newblock {Machine Learned Calabi--Yau Metrics and Curvature}.
\newblock 11 2022.

\bibitem{Gerdes:2022nzr}
Mathis Gerdes and Sven Krippendorf.
\newblock {CYJAX: A package for Calabi-Yau metrics with JAX}.
\newblock 11 2022.

\bibitem{Braun:2010vc}
Volker Braun.
\newblock {On Free Quotients of Complete Intersection Calabi-Yau Manifolds}.
\newblock {\em JHEP}, 04:005, 2011.

\bibitem{Braun:2017juz}
Andreas Braun, Andre Lukas, and Chuang Sun.
\newblock {Discrete Symmetries of Calabi\textendash{}Yau Hypersurfaces in Toric
  Four-Folds}.
\newblock {\em Commun. Math. Phys.}, 360(3):935--984, 2018.

\bibitem{Gray:2021kax}
James Gray and Juntao Wang.
\newblock {Free quotients of favorable Calabi-Yau manifolds}.
\newblock {\em JHEP}, 07:116, 2022.

\bibitem{Braun:2009mb}
Volker Braun.
\newblock {Three Generations on the Quintic Quotient}.
\newblock {\em JHEP}, 01:094, 2010.

\bibitem{Candelas:1987kf}
P.~Candelas, A.~M. Dale, C.~A. Lutken, and R.~Schimmrigk.
\newblock {Complete Intersection Calabi-Yau Manifolds}.
\newblock {\em Nucl. Phys. B}, 298:493, 1988.

\bibitem{Anderson:2015iia}
Lara~B. Anderson, Fabio Apruzzi, Xin Gao, James Gray, and Seung-Joo Lee.
\newblock {A new construction of Calabi\textendash{}Yau manifolds: Generalized
  CICYs}.
\newblock {\em Nucl. Phys. B}, 906:441--496, 2016.

\bibitem{Kreuzer:2000xy}
Maximilian Kreuzer and Harald Skarke.
\newblock {Complete classification of reflexive polyhedra in four-dimensions}.
\newblock {\em Adv. Theor. Math. Phys.}, 4:1209--1230, 2000.

\bibitem{Okonek1980}
Christian Okonek, Michael Schneider, and Heinz Spindler.
\newblock Vector bundles on complex projective spaces.
\newblock 1980.

\bibitem{Hubsch:1992nu}
Tristan Hubsch.
\newblock {\em {Calabi-Yau manifolds: A Bestiary for physicists}}.
\newblock World Scientific, Singapore, 1994.

\bibitem{Donagi:2004su}
Ron Donagi, Yang-Hui He, Burt~A. Ovrut, and Rene Reinbacher.
\newblock {Higgs doublets, split multiplets and heterotic SU(3)(C) x SU(2)(L) x
  U(1)(Y) spectra}.
\newblock {\em Phys. Lett. B}, 618:259--264, 2005.

\bibitem{Donagi:2004ub}
Ron Donagi, Yang-Hui He, Burt~A. Ovrut, and Rene Reinbacher.
\newblock {The Spectra of heterotic standard model vacua}.
\newblock {\em JHEP}, 06:070, 2005.

\bibitem{donaldson_2001}
S.K. Donaldson.
\newblock {Scalar Curvature and Projective Embeddings, I}.
\newblock {\em Journal of Differential Geometry}, 59(3):479 -- 522, 2001.

\bibitem{Griffiths:433962}
Phillip~A Griffiths and Joseph Harris.
\newblock {\em {Principles of algebraic geometry}}.
\newblock Wiley classics library. Wiley, New York, NY, 1994.

\bibitem{Wang}
Xiaowei Wang.
\newblock Canonical metrics on stable vector bundles.
\newblock {\em Communications in Analysis and Geometry}, 13, 03 2005.

\bibitem{Seyyedali}
Reza Seyyedali.
\newblock {Numerical Algorithm for Finding Balanced Metrics on Vector Bundles}.
\newblock {\em Asian Journal of Mathematics}, 13(3):311 -- 322, 2009.

\bibitem{James:1980yn}
F.~James.
\newblock {Monte Carlo Theory and Practice}.
\newblock {\em Rept. Prog. Phys.}, 43:1145, 1980.

\bibitem{Shiffman1999}
Bernard Shiffman and Steve Zelditch.
\newblock Distribution of zeros of random and quantum chaotic sections of
  positive line bundles.
\newblock {\em Communications in Mathematical Physics}, 200(3):661--683, Feb
  1999.

\bibitem{Shiffman2008}
Bernard Shiffman and Steve Zelditch.
\newblock Number variance of random zeros on complex manifolds.
\newblock {\em Geometric and Functional Analysis}, 18(4):1422--1475, Dec 2008.

\bibitem{Constantin:2018hvl}
Andrei Constantin and Andre Lukas.
\newblock {Formulae for Line Bundle Cohomology on Calabi-Yau Threefolds}.
\newblock {\em Fortsch. Phys.}, 67(12):1900084, 2019.

\bibitem{Anderson:2007nc}
Lara~B. Anderson, Yang-Hui He, and Andre Lukas.
\newblock {Heterotic Compactification, An Algorithmic Approach}.
\newblock {\em JHEP}, 07:049, 2007.

\bibitem{Anderson:2008uw}
Lara~B. Anderson, Yang-Hui He, and Andre Lukas.
\newblock {Monad Bundles in Heterotic String Compactifications}.
\newblock {\em JHEP}, 07:104, 2008.

\bibitem{Anderson:2008ex}
Lara~Briana Anderson.
\newblock {Heterotic and M-theory Compactifications for String Phenomenology}.
\newblock Other thesis, 8 2008.

\bibitem{Hoppe1984}
Hans~J{\"u}rgen Hoppe.
\newblock Generischer spaltungstyp und zweite chernklasse stabiler
  vektorraumb{\"u}ndel vom rang 4 auf p4.
\newblock {\em Mathematische Zeitschrift}, 187(3):345--360, Sep 1984.

\bibitem{Candelas:2007ac}
Philip Candelas, Xenia de~la Ossa, Yang-Hui He, and Balazs Szendroi.
\newblock {Triadophilia: A Special Corner in the Landscape}.
\newblock {\em Adv. Theor. Math. Phys.}, 12(2):429--473, 2008.

\end{thebibliography}
\end{document}